\documentclass[ aip,rsi, amsmath,amssymb, reprint, floatfix]{revtex4-1}

\usepackage{graphicx}% Include figure files
\usepackage{dcolumn}% Align table columns on decimal point
\usepackage{bm}% bold math
\usepackage{float}
\usepackage{changepage}
\usepackage[utf8]{inputenc}
\usepackage[T1]{fontenc}
\usepackage{mathptmx}
\PassOptionsToPackage{hyphens}{url}\usepackage{hyperref}
\usepackage{gensymb}
\usepackage[version=3]{mhchem}

\begin{document}

\preprint{AIP/123-QED}

\title[Extreme Ultraviolet-Excited Time-Resolved Luminescence Spectroscopy Using a High-Harmonic Generation Source]{Extreme Ultraviolet-Excited Time-Resolved Luminescence Spectroscopy Using an Ultrafast Table-Top High-Harmonic Generation Source}

\author{M. van der Geest}
\affiliation{ 
Advanced Research Center for Nanolithography, Science Park 106, 1098 XG, Amsterdam, The Netherlands
}%
\author{N. Sadegh}
\affiliation{ 
Advanced Research Center for Nanolithography, Science Park 106, 1098 XG, Amsterdam, The Netherlands
}%
\author{T.M. Meerwijk}
\affiliation{ 
Advanced Research Center for Nanolithography, Science Park 106, 1098 XG, Amsterdam, The Netherlands
}%
\author{E.I. Wooning}
\affiliation{ 
Advanced Research Center for Nanolithography, Science Park 106, 1098 XG, Amsterdam, The Netherlands
}%
\author{L. Wu}
\affiliation{ 
Advanced Research Center for Nanolithography, Science Park 106, 1098 XG, Amsterdam, The Netherlands
}%
\author{R. Bloem}
\affiliation{ 
Advanced Research Center for Nanolithography, Science Park 106, 1098 XG, Amsterdam, The Netherlands
}%
\author{S. Castellanos Ortega}
\affiliation{ 
Advanced Research Center for Nanolithography, Science Park 106, 1098 XG, Amsterdam, The Netherlands}
\author{A.M. Brouwer}
\affiliation{ 
Advanced Research Center for Nanolithography, Science Park 106, 1098 XG, Amsterdam, The Netherlands
}%
\affiliation{Van 't Hoff Institute for Molecular Sciences, University of Amsterdam, Science Park 904, 1098 XH, Amsterdam, Netherlands}
\author{P.M. Kraus}
\affiliation{ 
Advanced Research Center for Nanolithography, Science Park 106, 1098 XG, Amsterdam, The Netherlands
}%
\affiliation{LaserLaB, Department of Physics and Astronomy, Vrije Universiteit Amsterdam, De Boelelaan 1105, 1081 HV, Amsterdam, The Netherlands
}%
\affiliation{Author to whom correspondence should be addressed: \href{mailto:kraus@arcnl.nl}{kraus@arcnl.nl}}%

\date{\today}

\begin{abstract}
We present a table-top extreme ultraviolet (XUV) beamline for measuring time- and frequency-resolved XUV excited optical luminescence (XEOL) with additional femtosecond-resolution XUV transient absorption spectroscopy functionality. 
XUV pulses are generated via high-harmonic generation using a near-infrared pulse in a noble gas medium, and focused to excite luminescence from a solid sample.
The luminescence is collimated and guided into a streak camera, where its spectral components are temporally resolved with picosecond temporal resolution.
We time-resolve XUV excited luminescence and compare the results to luminescence decays excited at longer wavelengths for three different materials : (i) sodium salicylate, an often used XUV scintillator, (ii) fluorescent labeling molecule 4-carbazole  benzoic  acid (CB), and (iii) a zirconium metal oxo-cluster labeled with CB, which is a photoresist candidate for extreme-ultraviolet lithography. 
Our results establish time-resolved XEOL as a new technique to measure transient XUV-driven phenomena in solid-state samples, and identify decay mechanisms of molecules following XUV and soft-X-ray excitation.

\end{abstract}

\maketitle

\section{\label{sec:Intro}Introduction}
Photoinduced elementary molecular processes take place on ultrashort time-scales ranging from attoseconds to nanoseconds.\cite{Kraus2018a,Kraus2018,Geneaux2019}
High harmonic generation (HHG) in the extreme ultraviolet (XUV) from a fundamental infrared laser pulse is an important tool to study ultrafast processes, due to both the attainable ultrashort pulse durations as well as the short wavelengths of extreme ultraviolet (XUV) ($120$ to $10$ nm) and X-ray ($<10$ nm) pulses that are routinely generated through HHG, which enable photoionization,\cite{Drescher2002} photofragmentation \cite{Calegari2014} and XUV and X-ray absorption spectroscopies.\cite{Jager2017,Kaplan2018}
Transient XUV spectroscopy has been used to study fs oxidation state changes,\cite{Zhang2016} charge transfers,\cite{Kraus2015,Worner2017,Ryland2018} spin crossover,\cite{Zhang2019} and carrier dynamics in homogeneous materials\cite{Zurch2017a,Jager2017,Kaplan2018,Cushing2017} and across heterogeneous boundary layers.\cite{Cushing2020}\\
An emerging field of science and technology is extreme-ultraviolet ($13.5$ nm, EUV) nanolithography for printing integrated circuits, which is currently scaling up to high-volume manufacturing.\cite{Li2017}
The study of XUV-induced dynamics and XUV-initiated chemical reactions is therefore gaining interest.\cite{Haitjema2017a,Wu2019a,Sadegh2020} Most of the HHG studies mentioned above, however, excited dynamics at longer wavelengths and then probed the dynamics in the XUV.

Utilization of table-top HHG-based XUV setups as excitation sources for studying XUV-driven reactions has been somewhat limited by the inherently low HHG conversion efficiency and resulting poor photon fluxes.
Nevertheless, XUV-induced phenomena have been investigated, including electron-hole dynamics at a conical intersection,\cite{Timmers2014} few femtosecond electron-hole dynamics in ionized amino acids,\cite{Calegari2014} and ultrafast relaxation mechanisms of highly excited polycyclic organic molecules.\cite{Marciniak2015} 
All these experiments have in common that they measure gas-phase dynamics in small and medium-sized molecules, but their application to solid-state materials is more challenging.
X-ray (or XUV) excited optical luminescence (XEOL), a form of photoluminescence (PL), 
%Elango1983,Zimmerer2006,DeGrazia2007,Kirm2007,Vielhauer2008,
\cite{Vielhauer2009,Savchyn2012,Sokolov2012,Pustovarov2013,Belskii2016,Krzywinski2017,Chylii2019,Pankratov2020} is another technique that allows studying dynamics following XUV excitation.
XEOL was first applied for determining ppb-level lanthanide concentrations in host-crystals,\cite{Jaworowski1968} and was more recently reported to reproduce near edge X-ray absorption fine structures (NEXAFS) without having to transmit X-rays through the material.\cite{Soderholm1998,Rogalev2002}
Additional information can be gained by measuring the XEOL in a time-resolved manner (tr-XEOL), allowing one e.g. to determine excess carrier recombination lifetimes.\cite{OMalley2011}

XEOL experiments are readily carried out at suitable synchrotron endstations, such as was recently installed at the MAX IV synchrotron (Lund, Sweden) as well as the SUPERLUMI beamline at DESY, (Hamburg, Germany).\cite{Pankratov2020}
Some reported discoveries made at these endstations include investigations of the PL of silicon nanocrystals embedded in silicon dioxide\cite{Pankratov2011} and a thorough investigation of europium emission spectra in lanthanum trichloride host-crystals \cite{Savchyn2012}. 
So far, relatively few tr-XEOL studies have been reported using XUV from a table-top HHG source as the excitation source.\cite{Sekikawa2000, Sekikawa2002,Kirm2007}

Here, we present a versatile table-top setup that enables time-resolved luminescence spectroscopy with excitation wavelengths ranging from the XUV to IR. Compared to XUV pump-probe techniques that are challenged by the low conversion efficiency of HHG, the measurement of XEOL decay traces in a streak camera setup can easily be accomplished with the photon fluxes present in HHG pulses.
Our setup supports a large range of excitation wavelengths, ranging from near infrared (NIR) to deep ultraviolet (DUV) to XUV wavelengths.
XEOL and other accompanying fluorescence phenomena can be investigated in this setup in a $70$~ps to $500$~$\mu$s time range using a frequency-resolved streak camera system, with an ultimate temporal resolution of less than $1$~ps in the shortest time range.
An added pulse tunable from the NIR to DUV provides pump-probe functionality to allow for more extended material studies in the form of XUV transient absorption (XTAS) measurements and the setup is designed such that switching between XEOL and XTAS measurements can be done very quickly.\\
\\
For this paper, we studied and time-resolved the XUV excited luminescence of the scintillator sodium salicylate (NaSal), the fluorophore 4-carbazole  benzoic  acid (CB), and a zirconium metal oxo-cluster photoresist which incorporates this fluorophore.
This particular metal oxo-cluster has a core structure, consisting of six zirconium and four oxygen atoms plus four hydroxides, stabilized by twelve methacrylate (OMc) ligands (\ce{Zr6O4(OH)4(OMc)12} or ZrOMc).\cite{Kickelbick1997}
ZrOMc is a promising photoresist material for EUV lithography, and an important feature is its ability to switch solubility at low exposure dose after a photochemical reaction.
As this solubility switching is hard to observe directly, the fluorophore $4$-carbazole benzoic acid (CB) was attached to facilitate post-exposure spectroscopic measurements.\cite{Wu2019} 
Previously, it was reported that the introduction of CB required doubling the exposure dose for optimal solubility switching, known as dose-to-gel.\cite{Wu2019}
tr-XEOL measurements conducted with this setup provide additional insight into the reasons why the dose-to-gel increases two-fold, due to changing kinetics and energy loss to luminescence.

 \begin{figure*}[ht]
 \centering
     \includegraphics[width=\textwidth]{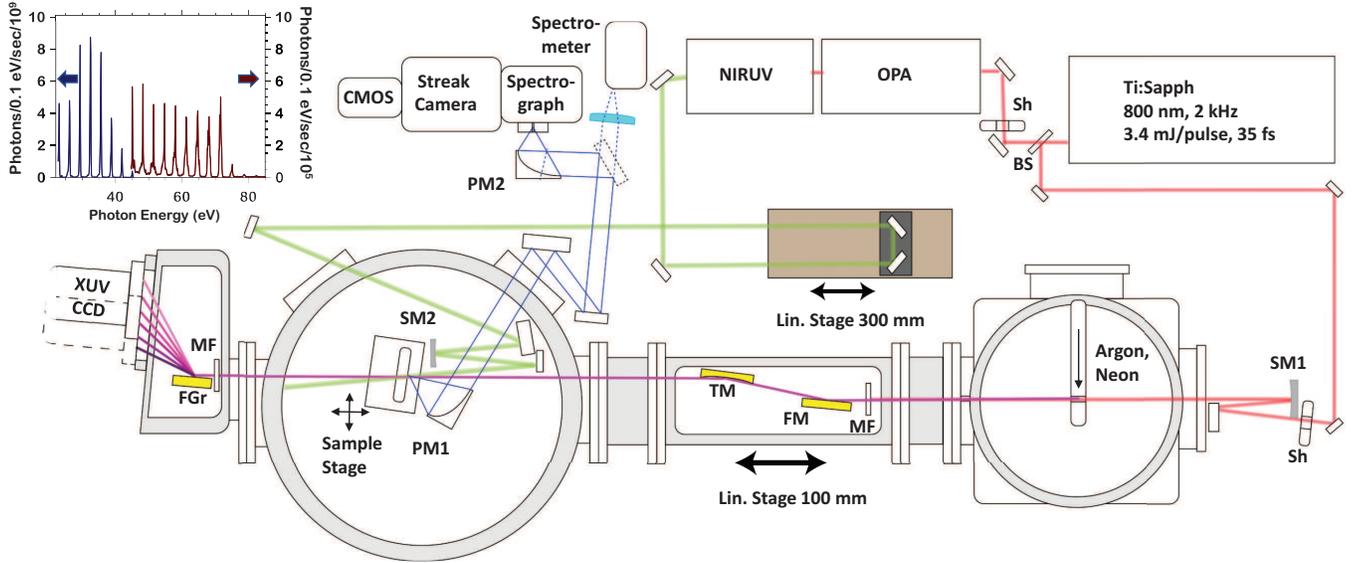}
         \caption{Schematic overview of the beamline. BS: $63:37$ beam splitter, with $37\%$ going toward the OPA. Sh: Shutter; SM$1$: Spherical mirror with f = $500$~mm; SM$2$: Spherical mirror with f = $250$~mm; MF: Metallic filter; FM: Flat mirror; TM: Toroidal mirror with f = $420$~mm; FGr: Flat-field XUV grating with f = $235$ mm; PM$1$: Parabolic mirror f = $50$~mm; PM$2$: Parabolic mirror f = $100$~mm. The focus chamber can be translated over a distance of $100$~mm, thus allowing the dose on sample to be varied if necessary. The XUV CCD can be rotated around its axis completely, giving a spectral range of $23$ to $48$~eV in the low energy position and $48$ to approximately $120$~eV in the high energy position. Inset: Calculated typical number of photons per 0.1 eV bin per second incident on the sample, using Eq. \ref{pulseenergy}, generated in argon (blue, left y-axis) and neon (red, right y-axis).}
     \label{Fig:Beamline}
 \end{figure*}
\section{Experimental Setup}\label{ExpOverview}

\subsection{Beamline Overview}

Fig.\ref{Fig:Beamline} shows a schematic overview of the table-top setup.
Pulses with $3.4$~mJ pulse energy, $35$~fs pulse duration, at a center wavelength of $800$~nm are generated by a titanium:sapphire laser system (Solstice ACE, Spectra-Physics) at a repetition rate of $2$~kHz. 
A beam splitter splits the pulse $63/37$, with $2.2$~mJ going towards the vacuum system containing the generating medium, while $1.2$~mJ pump an optical parametric amplifier (OPA) system (TOPAS-Prime, Light Conversion) with an added frequency upconversion module (NIRUVis, Light Conversion). 

The beamline consists of four vacuum chambers.
In front of the first chamber, the HHG chamber, the IR pulses are focused with a $500$~mm focal length spherical mirror into a gas cell with a diameter of $500$~$\mu$m and a length of $5$~mm, containing a low-pressure (between $20$ and $80$~mbar) noble gas, initiating the HHG process. 
The XUV source emits broadband spectra up to $45$~eV using argon (inset, Fig.\ref{Fig:Beamline}) and more than $70$~eV using neon.
For optimizing phase matching conditions, the gas pressure inside the gas cell can be adjusted with a mass-flow controller (IN-FLOW, Bronckhorst). An in-line setup for generating an $800+400$~nm two-color laser field to drive HHG can be inserted into the beam path. This arrangement consists of a BBO for frequency doubling, calcite plates for group delay compensation between the fundamental and second harmonic, and a subsequent $\lambda/2$ waveplate for 800 nm to rotate the polarization of the 800~nm pulses onto the polarization of the 400~nm pulses. Such two-color fields can improve flux and divergence,\cite{Kim2005,RoscamAbbing2019} and give rise to the emission of both odd and even harmonics, which improves spectral coverage for broadband XUV transient absorption measurements as demonstrated in section III.
The harmonics are then transmitted through a nickel-meshed $200$~nm thick metallic filter (Luxel), made of either aluminum or zirconium, to remove the fundamental.
A toroidal mirror (TM, $10$\degree~grazing incidence angle) reimages the harmonics on the sample.
The metallic filter and toroidal mirror are located in the second vacuum chamber, which itself is mounted onto a $100$~mm translation stage. This provides a method of increasing and decreasing the spot size and thus the fluence of the XUV pulses at a given sample position by changing the position of the XUV focus. Knife-edge measurements of the focal spot size yield a full width at half maximum (FWHM) of $43\pm7$~$\mu$m for the XUV. %accounting for long and short trajectory contributions,\cite{RoscamAbbing2019} 
This enables dose-dependent studies without the need of changing any parameters in the HHG process (such as driving intensity, pressure in gas cell), which would also affect the phase matching conditions and thus alter the spectral composition of the emitted harmonics.
The XUV light is then sent through a second $100$~nm thick aluminum filter (Luxel) to reduce background noise and is then spectrally resolved using an aberration-corrected flat-field XUV-grating ($1200$~l/mm, f=$235$~mm, grazing incidence angle $3$\degree, Hamamatsu) and measured using an XUV sensitized charge-coupled device (CCD) camera (GE $2048$ $512$ BI UV$1$, GreatEyes).
The CCD camera itself is mounted on an in-vacuum rotatable flange, which enables capturing a much larger spectral range without changing the grating angle from the specified, optimal angle of incidence.
At the optimal grazing incidence angle, the CCD camera captures a range of $23$ to $48$~eV in the low energy position, while capturing a range between $48$ and $120$~eV in the high energy position.
The XUV-grating is mounted on a two-axis rotational stage (Smaract) which provides additional flexibility in alignment and which energies are visible in a given stage position.

The OPA/NIR-UVVis system generates pulses between $230$ and $2500$~nm with approximately $300$~$\mu$J per pulse at IR wavelengths to approximately $10$~$\mu$J per pulse at DUV wavelengths. These pulses are focused on the sample using a $250$~mm spherical mirror to a spot size of and $153\pm2$~$\mu$m and temporally synchronized with the XUV pulses from HHG using a linear stage (M-$531$.DD$1$, PI) with long scanning range.
The output of the OPA functions both as a pulse for pump-probe experiments and as an excitation source to study carrier dynamics by photoluminescence.

Samples are positioned using a four dimensional stage (Smaract), which can be moved with a precision of $10$~nm and enables raster scanning if photobleaching of the fluorescence becomes an issue, as elaborated on in section~\ref{subsec:NaSal}.

The light collection beam path to the streak camera (C$10910$, Hamamatsu) consists of a $50$~mm focal length parabolic mirror (PM$1$), capturing up to $6\%$ of emitted light at a near normal angle. 
The luminescence is refocused using a $100$~mm focal length parabolic mirror (PM$2$) onto the slit of a Czerny-Turner scheme-based spectrograph (SP-$2300$i, Princeton Instruments) mounted in front of the streak system, containing three different gratings optimized for different spectral ranges.
The streak camera operates using two electronic sweep units to cover different time ranges.
The slow sweep unit is synchronized with the amplifier ($2$~kHz) and is utilized in measurements in the range of $500$~$\mu$s to $2$~ns, while the synchroscan unit is synchronized with the seed laser output ($84$~MHz) and is used for time windows shorter than $2$~ns, down to $70$~ps.
The time-delayed electrons are thus spatially spread out by the electric field before they hit a multi-channel plate (MCP) coupled to a phosphor screen.
The signal from the phosphor screen is recorded by a complementary metal-oxide-semiconductor (CMOS) camera (ORCA-FLASH$4.0$, Hamamatsu).
Additionally, a removable mirror combined with a $150$~mm focal length lens are included with a spectrometer (Maya~$2000$ Pro, Ocean Optics) in the light collection beam path, as can also be seen in Fig.\ref{Fig:Beamline}. This enables measuring the emission spectrum over time and continuous exposure.
Due to the low photon fluxes of HHG, XUV-excited time-resolved luminescence traces of the investigated samples can have low count levels, which requires careful optimization of the experimental setup. To optimize data acquisition, a high quantum efficiency (QE) alignment target is used for optimizing streak measurements, consisting of sodium salicylate (NaSal, $99\%$~purity, Sigma-Aldrich) dropcasted from a saturated methanol solution on a $1.2x1.2$ mm$^2$ quartz substrate. 
NaSal forms a coarse white thin film which emits bright blue, broad luminescence at $428$~nm ($2.9$~eV) upon excitation.
Electronic trigger drift correction during long measurement campaigns is implemented by sending an attenuated fs pulse into the streak camera and post correcting a series of measurements on the instrument-response-function (IRF) limited signal generated by this pulse.\cite{Yamanoi2010}

\subsection{XUV Flux Estimate}
The XUV pulse energy $E_{\text{pulse}}$ at the sample position was estimated to be approximately $0.50$~nJ/pulse, which is equivalent to a fluence of approximately $26$~$\mu$J/cm$^2$ when generating in argon.\cite{Manschwetus2016} 
The estimated flux in neon is much lower at $20$~pJ/pulse, due to its higher ionization potential, but also yields a higher cutoff energy.
For ease of comparison, the number of photons is converted to photons/$0.1$~eV/second in the inset of Fig.\ref{Fig:Beamline}.
The XUV pulse energy is calculated from a representative CCD measurement by using Eq. \ref{pulseenergy}.

 \begin{equation}\label{pulseenergy}
 \centering
  E_{\text{pulse}}\approx \frac{E_{\text{spec}}(eV)}{A_{\text{tot}}(eV)\eta_{\text{grat}}(eV)\eta_{\text{eh}}(eV)\Phi_{\text{CCD}}(eV)\text{SN}} .
\end{equation}

In Eq. \ref{pulseenergy} a recorded XUV spectrum, normalized to the number of integrated pulses, ($E_\mathrm{spec}(eV)$) is divided by the estimated total absorption ($A_\mathrm{tot}(eV)$) of a $100$~nm aluminum filter with an estimated $8$~nm aluminum oxide layer ($4$~nm on the front- and backsides)\cite{Campbell1999} as well as the polarization dependent grating efficiency $\eta_{\text{grat}}(eV)$,\cite{Frassetto2010} and the CCD electron-hole pair creation efficiency ($\eta_{\text{eh}}(eV)$).\cite{Scholze1998} 
The CCD quantum efficiency ($\Phi_{\text{CCD}}(eV)$) and the acquisition-setting dependent gain factor of the CCD ($\text{SN}$) were both provided by the supplier.

\begin{figure}[ht]
    \includegraphics[width=0.5\textwidth]{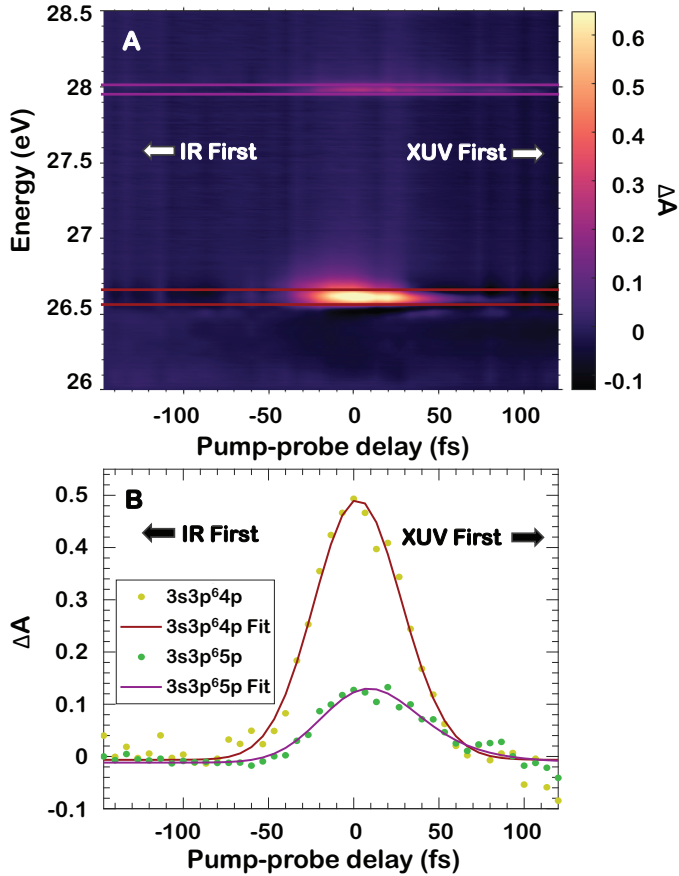}
    \caption{Argon auto-ionization states corresponding to the $3s^23p^6$~to~$3s3p^64p$~($4$p) and $3s^23p^6$~to~$3s3p^65p$~($5$p) transitions are perturbed by a non-collinear IR field centered at $860$~nm, which decreases the re-emission of the XUV light.
    (A) XUV absorption change as function of pump-probe delay.
    (B) Averaged line-outs of both resonances across the pump-probe delay, showing FWHM's of the temporal instrument response function of $58\pm9$ and $57\pm8$~fs, for the $4$p and $5$p transitions, respectively.}
    \label{fig:ArgonResonances}
\end{figure}
\section{\label{sec:XTAS}XUV Transient Absorption}

The XTAS setup is installed in such a manner that switching between XTAS and XEOL experiments can be done very quickly.
The temporal and spectral resolutions are important in XTAS experiments, as they allow for \textit{e.g.} temporal drift corrections and observing absorption features in high resolution.
In order to determine the temporal and spectral resolution of XTAS, a second gas cell, with a length of $5$~mm and diameter of $500$~$\mu$m is moved in the focus of the pump-probe overlap using a secondary, linear stage.
This gas cell is filled with argon up to a pressure of approximately 10-20 mbar, resulting in significant XUV absorption.
Due to the excitation of argon auto-ionization states, several sharp lines will appear in the spectrum between $26.5$ and $29$ eV where XUV light is re-emitted.\cite{Wang2013,Madden1969}
Perturbing the XUV-excited polarization decay of auto-ionization states with a temporally synchronized IR pulse will decrease the XUV re-emission of these auto-ionization states.\cite{Wang2010}

In Fig. \ref{fig:ArgonResonances}A, a spectrogram can be seen depicting the difference in absorption ($\Delta$A) observed across the pump-probe delay of the $860$~nm probe-pulse and the XUV broadband pump between $26$ and $28.5$~eV.
The auto-ionization states that can be clearly observed are the $3s^23p^6$~to~$3s3p^64p$ (abbreviated as $4$p) and $3s^23p^6$~to~$3s3p^65p$ ($5$p) transitions, centered at $26.6$ and $28.0$~eV, respectively.
These can be used for post-correcting long XTAS measurements due to their characteristic temporal shape and sharpness.
We then fit a Gaussian representing the temporal instrument response function convoluted with an exponential with decay constants given by the lifetime of the states ($8.2$~fs for the $4$p and $23.3$~fs for the $5$p transitions, respectively\cite{Wang2010}) to the averaged $\Delta$A in Fig. \ref{fig:ArgonResonances}B. We find instrument response times of $58\pm9$~fs and $57\pm8$~fs for the $4$p and $5$p transitions, respectively. The IRF is limited by the pulse duration coming out of the OPA, which introduces stretching of the pump pulse up to 60 fs as measured through frequency resolved optical gating.

As the spectral bandwidths of the argon auto-ionization states have been determined with very high precision,\cite{Madden1969} their observed bandwidths on the CCD spectrometer $\Gamma_\text{exp}$ can be used to determine the spectral resolution of the XUV spectrometer in this spectral region.\cite{Wang2013}
The energy resolution $\Gamma_\text{res}$ of the XUV spectrometer was determined through Eq.\ref{equ:res}.

\begin{equation}
    \Gamma_\text{res}=\sqrt{\Gamma_\text{exp}^2-\Gamma^2}.
    \label{equ:res}
\end{equation}

$\Gamma_\text{res}$ was determined to be $44.9\pm0.9$~meV around $28$~eV using the known FWHM $\Gamma$ of the $5$p auto-ionization state, $26.2$~meV,\cite{Wang2013} and the measured FWHM $\Gamma_\text{exp} = 52$~meV.
Neon has a similar series of auto-ionizing states starting at $45.6$~ eV.\cite{Codling1967} 
These auto-ionizing states were found to all have an approximate FWHM of $\Gamma_\text{exp}=80$~meV.
Using the same method described in this section and utilizing Eq. \ref{equ:res}, the energy resolution around these states was determined to be $78\pm1$~meV. 
At higher energies, even better resolving powers are expected, as the grating is optimized for the energy range $60-120$ eV.

\begin{figure*}[ht]
\centering
\includegraphics[width=\textwidth]{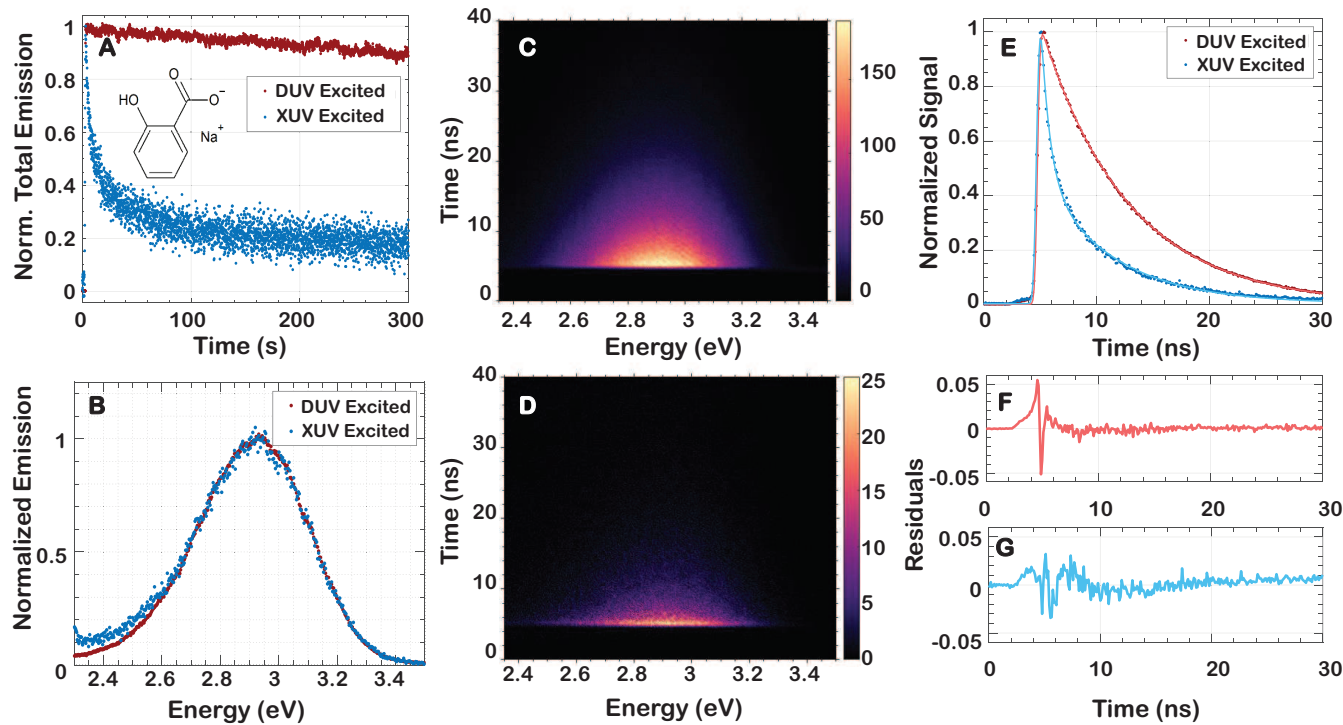}
\caption{Comparison of NaSal excited by pulsed DUV and XUV radiation. (A) Sodium salicylate (inset) emission intensity over time under continuous pulsed DUV and XUV excitation. (B) Emission spectrum upon DUV and XUV excitation (C) Streak trace of DUV excited NaSal. (D) Streak trace of XUV excited NaSal (E) Comparison and fit of time-resolved emissions of DUV and XUV. (F,G) Residuals of the DUV and XUV fit in (E), respectively.}
\label{Fig:sodiumstreak}
\end{figure*}
\section{\label{sec:STREAK}XUV Excited Optical Luminescence}
\subsection{\label{subsec:NaSal}XEOL Behavior of Sodium Salicylate}

NaSal (Fig.\ref{Fig:sodiumstreak}A) was selected as a benchmark for characterizing the beamline, as it has a high, near constant QE across the XUV range up to the carbon K-edge at $280$~eV, and comparative literature is available.\cite{Samson1974,Moine2007}
As can be seen in Fig.\ref{Fig:sodiumstreak}A, the emission of the NaSal XEOL over time was found to drop by about $66\%$ within $15$~seconds at $26$~$\mu$J/cm$^2$, delivering a total dose of $720$~mJ/cm$^2$ during that time.
This contrasts very strongly with the vacuum emission over time when excited by $266$~nm pulses from the OPA with a single-shot fluence of $450$~$\mu$J/cm$^2$, where the fluence only drops by about $10$\% over the course of five minutes, although a much higher dose was delivered ($13.5$~J/cm$^2$ in $15$ sec, $270$~J/cm$^2$ in $5$ min). 
The much shorter absorption length of XUV compared to DUV is responsible for this contrasting behavior.
The excited state concentration at the surface is so high that bleaching processes take place in much shorter time-scales.
During the XUV photobleaching we observe that the emission spectrum does not change, meaning that the emitting state or species is likely not changing either.
The emission spectrum of XUV and DUV excitation is also nearly the same, as can be seen in Fig. \ref{Fig:sodiumstreak}B, hinting that sample decomposition rather than reaction with other compounds is the reason for the photobleaching.
Therefore, the NaSal sample was raster scanned by moving to a new spot $80$~$\mu$m away every $10$~seconds to ensure a relatively constant emission intensity.
Given the size of the sample at $1.44$~cm$^2$ , this means that the sample has to be replaced after approximately $6$~hours of continuous raster scanning, although NaSal samples were replaced more often than that for convenience.

Two streak traces were then measured for DUV and for XUV excitation, respectively in Fig.\ref{Fig:sodiumstreak}C and Fig.\ref{Fig:sodiumstreak}D.
A $280$~nm longpass filter was used to filter out the excitation pulse.
A mono-exponential fit ($R^2=0.998$, $\chi^2=0.16$) of the time-resolved signal intensity trace for NaSal excited by $34$~mJ/cm$^2$ $266$~nm pulses (Fig.\ref{Fig:sodiumstreak}E, red curve) yields $\tau_1=7.59\pm0.03$~ns.
%The residuals in Fig.\ref{Fig:sodiumstreak}F show some structure, but this is related to an electronic artifact that appears as a small, pulse signal right before the start of the decay.
The residuals in Fig.\ref{Fig:sodiumstreak}F show no significant structure except for before the start of the decay. This is related to an electronic artifact in the streak camera.
Solid sodium salicylate has reported mono-exponential decay constants of between $6.5$~ns and $8.5$~ns under DUV excitation, depending on sample preparation, and $\tau_1$ is thus in good agreement with previously reported values.\cite{Sigmond1966,Waynant1971,Baker1987,Kolasinski2002}

The time-resolved emission changes significantly when the same NaSal sample is excited by XUV pulses, as can be seen in the blue curve in Fig.\ref{Fig:sodiumstreak}E, becoming both faster and bi-exponential in nature.
Fitting the signal intensity trace in Fig.\ref{Fig:sodiumstreak}E with a bi-exponential decay results in a good fit (Fig.\ref{Fig:sodiumstreak}G, R$^2=0.992$,  $\chi^2=0.64$), yielding values for $\tau_1$ and $\tau_2$ of $0.77(\pm0.04)$~ns and  $6.28\pm0.14$~ns with proportionality constants $\alpha_1=0.64$ and $\alpha_2=0.36$, respectively.
This apparent change in decay when excited with $>15$ eV pulsed sources is attributed to the rapid quenching of excited states by ions and excited states located on neighboring molecules.\cite{Brocklehurst1997}
The values of $\tau_1$ and $\tau_2$ are dependent on the incoming photon energy: $\tau_1$ is found to be shorter with higher photon energy, while $\tau_2$ behaves oppositely.\cite{Baker1987,Brocklehurst1992a}
No $\tau_1$ and $\tau_2$ emission wavelength dependence was observed, with decay constants being the same within error bars as when the  emission spectrum is integrated around $3.1$, $2.9$ and $2.6$ eV.
No significant flux dependence of $\tau_1$ and $\tau_2$ was observed in measurements with XUV fluxes between $5$ and $30$~$\mu$J/cm$^2$.
\begin{figure}[ht]
\centering
\includegraphics[width=0.5\textwidth]{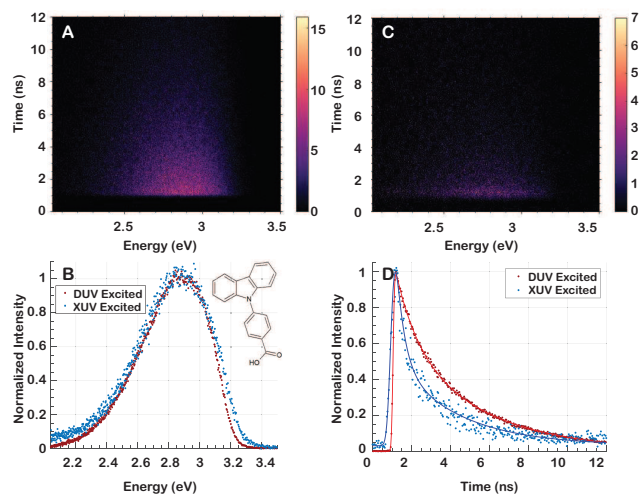}
\caption{Comparison of CB emission spectra, excited by pulsed $266$~nm DUV and XUV radiation. (A) Streak trace of CB under $266$~nm excitation. (B) Emission Spectra of CB under DUV (red curve) and XUV (blue curve) excitation. (C) Streak trace of CB under XUV excitation. (D) Decay of CB under DUV (red curve) and XUV (blue curve) excitation.}
\label{Fig:carbstreak}
\end{figure}
\subsection{\label{subsec:CB}Characterizing 4-Carbazole Benzoic Acid}

4-Carbazole  benzoic  acid (CB, structure inset Fig.\ref{Fig:carbstreak}B) is a fluorophore that can be used to label and monitor decay kinetics in ZrOMc photoresists and aid in characterizing condensation reactions of ZrOMc.
The streak trace of CB, excited by $266$~nm DUV pulses, can be seen in Fig.\ref{Fig:carbstreak}A.
A $280$~nm longpass filter was used to prevent artifacts related to the second order of the grating at $2.32$~eV.
Similar to NaSal, the emission spectrum of CB does not change significantly between XUV and DUV excitation, although some broadening occurs, as can be seen in Fig.\ref{Fig:carbstreak}B.
A best fit (R$^2=0.9993$, $\chi^2=0.137$) of the CB $266$~nm excited decay (Fig. \ref{Fig:carbstreak}D, red curve) yields a short component of $\tau_1=1.11\pm0.06$~ns and a long component of $\tau_2= 4.21\pm0.7$~ns with $\alpha_1=0.37$ and $\alpha_2=0.63$, respectively, which within error bars do not depend on which emission wavelength is considered.
This bi-exponential behavior is attributed to quenching due to overlap between $\pi$-orbitals of neighboring aromatic carbazole groups ($\pi$-stacking), which causes a deviation from a mono-exponential luminescence decay.\cite{Xue2015,Wang2016,Liu2020}
The decay becomes faster, and remains bi-exponential, in the case of XUV excitation, as can be seen in the blue curve of Fig. \ref{Fig:carbstreak}D.
Another instrument-related difference between DUV and XUV excitation can be seen in the apparent longer rising edge of the decay, which is caused by opening the entrance slit of the streak camera an extra 20 $\mu$m to increase the number of incoming photons.
A best fit (R$^2=0.979$, $\chi^2=2.2$) gives a short component of
$\tau_1=0.4\pm0.1$~ns and a long component $\tau_2=3.2\pm0.2$~ns combined with an almost inverse ratio of $\alpha_1=0.66$ and $\alpha_2=0.34$.
The value of these constants again does not depend on the emission wavelength under consideration.
This behavior of faster bi-exponential decay when excited with pulsed XUV radiation is observed in many solid aromatics and is rationalized to originate from additional quenching by a high density of excited state dipoles interacting with each other, while partially stabilized by aromatic chains.\cite{Brocklehurst1992a}

\subsection{\label{subsec:ZrOMc}Luminescence Dynamics of a Tagged Photoresist}
\begin{figure}[ht]
\centering
\includegraphics[width=0.5\textwidth]{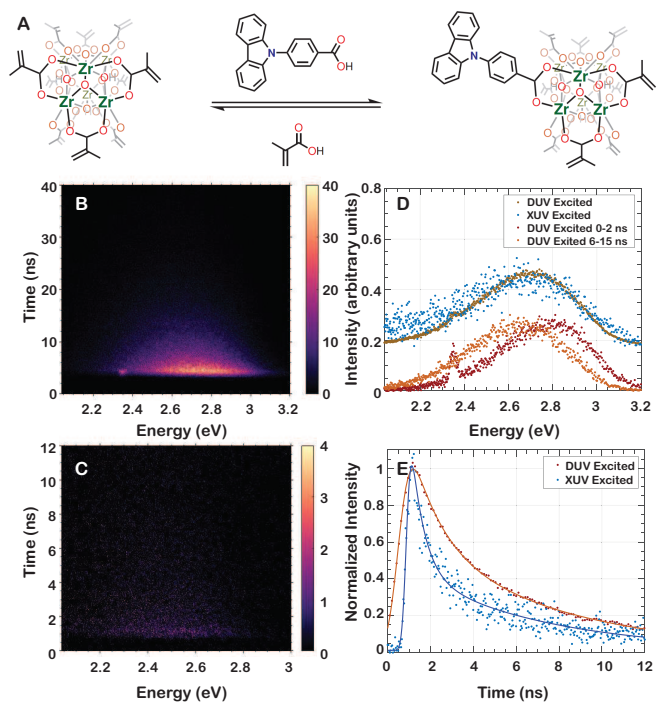}
\caption{Comparison of ZrOMc-CB tr-XEOL. (A) Equilibrium chemical reaction of ZrOMc and CB forming ZrOMc-CB. (B) Streak trace of ZrOMc-CB under $266$~nm DUV excitation. (C) Streak trace of ZrOMc-CB under XUV excitation. (D) Emission spectra showing a comparison between DUV (brown) and XUV (blue) emission spectra and the red shift in time (lower plots, with 2nd order grating artifact at $2.32$~eV). (E) First $12$~ns of the decay of ZrOMc-CB under DUV (red) and XUV (blue) excitation.}
\label{Fig:ZrStreak}
\end{figure}
Finally, a zirconium metal oxo-cluster labeled with the CB fluorophore was investigated.
As mentioned, ZrOMc-CB is formed by exchanging one of the OMc ligands with CB in a ligand-exchange reaction, which makes the this zirconium metal oxo-cluster luminescent and hence enables characterizing decay kinetics through luminescence.
The chemical structure of ZrOMc as well as the ligand-exchange reaction to form ZrOMc-CB can be viewed in Fig.\ref{Fig:ZrStreak}A.
The reaction scheme is similar to an earlier published reaction scheme.\cite{Wu2020}
When excited by $266$~nm DUV pulses, the streak trace differs significantly from CB alone, as it is both red-shifted and decays on a longer time-scale.
Fig.\ref{Fig:ZrStreak}B shows the DUV excited streak spectrum, with a $2$nd order grating artifact at $2.32$~eV.\\
A strong redshift after the first $4$~ns is also visible, likely caused by the emission of free CB which decays much faster than the coordinated fluorophore.
This is supported by the fact that the early emission spectrum of ZrOMc-CB in the first $2$~ns in Fig.\ref{Fig:ZrStreak}D agrees well with the emission spectrum of free CB in Fig.\ref{Fig:carbstreak}B.\\
The emission excited by pulsed XUV radiation was not very intense, but still measurable as can be seen in Fig.\ref{Fig:ZrStreak}D.
Despite the lower signal-to-noise ratio, as can be seen in Fig.\ref{Fig:ZrStreak}D top curves, no change of the emission spectrum is observed, with both free CB and the ZrOMc-CB itself contributing to the emission.
The bottom two curves in Fig.\ref{Fig:ZrStreak}D depict the large red shift in time visible in figure Fig.\ref{Fig:ZrStreak}B from $2.9$~eV to around $2.65$~eV.
Fitting the DUV decay (red curve) in Fig.\ref{Fig:ZrStreak}E in time yields $\tau_1=7.08\pm0.27$~ns and  $\tau_2=1.69\pm0.28$~ns (R$^2=0.996$, $\chi^2=0.4$) with proportionality constants $\alpha_1=0.61$ and $\alpha_2=0.39$.
The XUV decay (blue curve) has a lower signal to noise ratio, but is still significantly faster as is the case with NaSal and CB itself.
What is also apparent is the much longer rising edge of the decay in the case of DUV excitation. 
The longer apparent decay necessitated measuring in a longer time-range in the streak camera. This results in a longer IRF, which was shifted so that XUV and DUV excited decays both start at the same moment in time.
Nevertheless, the long component is within error bars of the long component of the DUV excited decay, yielding $\tau_1=6.26\pm1.44$~ns and  $\tau_2=0.77\pm0.14$~ns (R$^2=0.94$, $\chi^2=2.8$) with again near inverse proportionality constants $\alpha_1=0.32$ and $\alpha_2=0.68$.
One conclusion to be drawn is that the fact that the XUV excited emission could be observed at all means that some of the absorbed energy is re-emitted instead of utilized in solubility switching, thus providing an additional reason for the two-fold increase in dose requirements.\cite{Wu2019}
The faster decay does suggest that the luminescence is quenched more than during DUV excitation, possibly by the aromatic groups as in the case of CB.

\section{Conclusion}
In conclusion, we have reported on a table-top XUV beamline capable of measuring luminescence excited by visible, DUV and XUV sources in a time- and frequency resolved manner using an integrated streak camera.
By measuring both a high QE scintillator, a fluorophore and a photoresist material tagged with the fluorophore we also demonstrated its versatility in measuring different types of luminescent materials.
We demonstrated that the poor XUV photochemical reactivity performance of the ZrOMc-CB compared to the unsubstituted ZrOMc can be attributed to the fact that the substituted resist itself converts part of the energy to emitted photons.
The addition of a high temporal resolution, IR to DUV pump/XUV probe allows for XTAS measurements to get a more complete view of molecular and electronic dynamics on a fs to ns time-scale in relevant samples and, if necessary, XEOL and XTAS measurements can be run simultaneously.
This beamline can be used to study the dynamics in a wide variety of relevant solid materials including novel scintillators, $2$D materials, other photoresists, solar cell materials and strongly correlated materials. The XUV excitation is reminiscent of the photoactivation and radiation damage that such materials experience in modern-day nanolithography or even in space, where solar power panels may become more prevalent due to possibly higher collection efficiencies.
We expect that time-resolved XEOL will become an important tool to study charge carrier and exciton dynamics in solid-state materials excited by XUV pulses. In particular, XUV pulses typically have the shortest absorption lengths of all wavelengths which leads to a high carrier concentration in the first few nm near the surface.\cite{Henke1993} This makes XUV exited time-resolved luminescence ideally suited to study XUV-induced surface recombination mechanisms and kinetics.

\section*{Data availability}
The data that support the findings of this study are available from the corresponding author upon reasonable request.

\begin{acknowledgments}
This work was carried out at the Advanced Research Center for Nanolithography (ARCNL), a public-private partnership of the University of Amsterdam (UvA), the Vrije Universiteit Amsterdam (VU), the Netherlands Organisation for Scientific Research (NWO), and the semiconductor equipment manufacturer ASML.
We would like to thank Reinout Jaarsma and Sander van Leeuwen for technical support, Dirk-Jan Spaanderman for solving many design issues and Filippo Campi for providing valuable comments.
We also want to extend our thanks to the cleanroom support staff at AMOLF for help in sample preparation.
We furthermore wish to thank the workshop as well as the mechanical, electronic and software engineering departments at AMOLF and ARCNL for the construction and implementation of the setup.  P.M.K. acknowledges support from NWO Veni grant 016.Veni.192.254.
\end{acknowledgments}

\appendix

\section*{References}
\nocite{}


\begin{thebibliography}{63}%

\makeatletter
\providecommand \@ifxundefined [1]{%
 \@ifx{#1\undefined}
}%
\providecommand \@ifnum [1]{%
 \ifnum #1\expandafter \@firstoftwo
 \else \expandafter \@secondoftwo
 \fi
}%
\providecommand \@ifx [1]{%
 \ifx #1\expandafter \@firstoftwo
 \else \expandafter \@secondoftwo
 \fi
}%
\providecommand \natexlab [1]{#1}%
\providecommand \enquote  [1]{``#1''}%
\providecommand \bibnamefont  [1]{#1}%
\providecommand \bibfnamefont [1]{#1}%
\providecommand \citenamefont [1]{#1}%
\providecommand \href@noop [0]{\@secondoftwo}%
\providecommand \href [0]{\begingroup \@sanitize@url \@href}%
\providecommand \@href[1]{\@@startlink{#1}\@@href}%
\providecommand \@@href[1]{\endgroup#1\@@endlink}%
\providecommand \@sanitize@url [0]{\catcode `\\12\catcode `\$12\catcode
  `\&12\catcode `\#12\catcode `\^12\catcode `\_12\catcode `\%12\relax}%
\providecommand \@@startlink[1]{}%
\providecommand \@@endlink[0]{}%
\providecommand \url  [0]{\begingroup\@sanitize@url \@url }%
\providecommand \@url [1]{\endgroup\@href {#1}{\urlprefix }}%
\providecommand \urlprefix  [0]{URL }%
\providecommand \Eprint [0]{\href }%
\providecommand \doibase [0]{http://dx.doi.org/}%
\providecommand \selectlanguage [0]{\@gobble}%
\providecommand \bibinfo  [0]{\@secondoftwo}%
\providecommand \bibfield  [0]{\@secondoftwo}%
\providecommand \translation [1]{[#1]}%
\providecommand \BibitemOpen [0]{}%
\providecommand \bibitemStop [0]{}%
\providecommand \bibitemNoStop [0]{.\EOS\space}%
\providecommand \EOS [0]{\spacefactor3000\relax}%
\providecommand \BibitemShut  [1]{\csname bibitem#1\endcsname}%
\let\auto@bib@innerbib\@empty
%</preamble>
\bibitem [{\citenamefont {Kraus}\ \emph {et~al.}(2018)\citenamefont {Kraus},
  \citenamefont {Z{\"{u}}rch}, \citenamefont {Cushing}, \citenamefont
  {Neumark},\ and\ \citenamefont {Leone}}]{Kraus2018a}%
  \BibitemOpen
  \bibfield  {author} {\bibinfo {author} {\bibfnamefont {P.~M.}\ \bibnamefont
  {Kraus}}, \bibinfo {author} {\bibfnamefont {M.}~\bibnamefont {Z{\"{u}}rch}},
  \bibinfo {author} {\bibfnamefont {S.~K.}\ \bibnamefont {Cushing}}, \bibinfo
  {author} {\bibfnamefont {D.~M.}\ \bibnamefont {Neumark}}, \ and\ \bibinfo
  {author} {\bibfnamefont {S.~R.}\ \bibnamefont {Leone}},\ }\href {\doibase
  10.1038/s41570-018-0008-8} {\bibfield  {journal} {\bibinfo  {journal} {Nature
  Reviews Chemistry}\ }\textbf {\bibinfo {volume} {2}},\ \bibinfo {pages} {82}
  (\bibinfo {year} {2018})}\BibitemShut {NoStop}%
\bibitem [{\citenamefont {Kraus}\ and\ \citenamefont
  {W{\"{o}}rner}(2018)}]{Kraus2018}%
  \BibitemOpen
  \bibfield  {author} {\bibinfo {author} {\bibfnamefont {P.~M.}\ \bibnamefont
  {Kraus}}\ and\ \bibinfo {author} {\bibfnamefont {H.~J.}\ \bibnamefont
  {W{\"{o}}rner}},\ }\href {\doibase 10.1002/anie.201702759} {\bibfield
  {journal} {\bibinfo  {journal} {Angewandte Chemie - Int. Edition}}\
  }\textbf {\bibinfo {volume} {57}},\ \bibinfo {pages} {5228} (\bibinfo {year}
  {2018})\BibitemShut {NoStop}%
\bibitem [{\citenamefont {Geneaux}\ \emph {et~al.}(2019)\citenamefont
  {Geneaux}, \citenamefont {Marroux}, \citenamefont {Guggenmos}, \citenamefont
  {Neumark},\ and\ \citenamefont {Leone}}]{Geneaux2019}%
  \BibitemOpen
  \bibfield  {author} {\bibinfo {author} {\bibfnamefont {R.}~\bibnamefont
  {Geneaux}}, \bibinfo {author} {\bibfnamefont {H.~J.}\ \bibnamefont
  {Marroux}}, \bibinfo {author} {\bibfnamefont {A.}~\bibnamefont {Guggenmos}},
  \bibinfo {author} {\bibfnamefont {D.~M.}\ \bibnamefont {Neumark}}, \ and\
  \bibinfo {author} {\bibfnamefont {S.~R.}\ \bibnamefont {Leone}},\ }\href
  {\doibase 10.1098/rsta.2017.0463} {\bibfield  {journal} {\bibinfo  {journal}
  {Philosophical Transactions of the Royal Society A}\ }\textbf {\bibinfo
  {volume} {377}} (\bibinfo {year} {2019}),\
  10.1098}\BibitemShut {NoStop}%
\bibitem [{\citenamefont {Drescher}\ \emph {et~al.}(2002)\citenamefont
  {Drescher}, \citenamefont {Hentschel}, \citenamefont {Kienberger},
  \citenamefont {Uiberacker}, \citenamefont {Yakovlev}, \citenamefont
  {Scrinzi}, \citenamefont {Westerwalbesloh}, \citenamefont {Kleineberg},
  \citenamefont {Heinzmann},\ and\ \citenamefont {Krausz}}]{Drescher2002}%
  \BibitemOpen
  \bibfield  {author} {\bibinfo {author} {\bibfnamefont {M.}~\bibnamefont
  {Drescher}}, \bibinfo {author} {\bibfnamefont {M.}~\bibnamefont {Hentschel}},
  \bibinfo {author} {\bibfnamefont {R.}~\bibnamefont {Kienberger}}, \bibinfo
  {author} {\bibfnamefont {M.}~\bibnamefont {Uiberacker}}, \bibinfo {author}
  {\bibfnamefont {V.}~\bibnamefont {Yakovlev}}, \bibinfo {author}
  {\bibfnamefont {A.}~\bibnamefont {Scrinzi}}, \bibinfo {author} {\bibfnamefont
  {T.}~\bibnamefont {Westerwalbesloh}}, \bibinfo {author} {\bibfnamefont
  {U.}~\bibnamefont {Kleineberg}}, \bibinfo {author} {\bibfnamefont
  {U.}~\bibnamefont {Heinzmann}}, \ and\ \bibinfo {author} {\bibfnamefont
  {F.}~\bibnamefont {Krausz}},\ }\href {\doibase 10.1038/nature01143}
  {\bibfield  {journal} {\bibinfo  {journal} {Nature}\ }\textbf {\bibinfo
  {volume} {419}},\ \bibinfo {pages} {803} (\bibinfo {year}
  {2002})}\BibitemShut {NoStop}%
\bibitem [{\citenamefont {Calegari}\ \emph {et~al.}(2014)\citenamefont
  {Calegari}, \citenamefont {Ayuso}, \citenamefont {Trabattoni}, \citenamefont
  {Belshaw}, \citenamefont {{De Camillis}}, \citenamefont {Anumula},
  \citenamefont {Frassetto}, \citenamefont {Poletto}, \citenamefont {Palacios},
  \citenamefont {Decleva}, \citenamefont {Greenwood}, \citenamefont
  {Mart{\'{i}}n},\ and\ \citenamefont {Nisoli}}]{Calegari2014}%
  \BibitemOpen
  \bibfield  {author} {\bibinfo {author} {\bibfnamefont {F.}~\bibnamefont
  {Calegari}}, \bibinfo {author} {\bibfnamefont {D.}~\bibnamefont {Ayuso}},
  \bibinfo {author} {\bibfnamefont {A.}~\bibnamefont {Trabattoni}}, \bibinfo
  {author} {\bibfnamefont {L.}~\bibnamefont {Belshaw}}, \bibinfo {author}
  {\bibfnamefont {S.}~\bibnamefont {{De Camillis}}}, \bibinfo {author}
  {\bibfnamefont {S.}~\bibnamefont {Anumula}}, \bibinfo {author} {\bibfnamefont
  {F.}~\bibnamefont {Frassetto}}, \bibinfo {author} {\bibfnamefont
  {L.}~\bibnamefont {Poletto}}, \bibinfo {author} {\bibfnamefont
  {A.}~\bibnamefont {Palacios}}, \bibinfo {author} {\bibfnamefont
  {P.}~\bibnamefont {Decleva}}, \bibinfo {author} {\bibfnamefont {J.~B.}\
  \bibnamefont {Greenwood}}, \bibinfo {author} {\bibfnamefont {F.}~\bibnamefont
  {Mart{\'{i}}n}}, \ and\ \bibinfo {author} {\bibfnamefont {M.}~\bibnamefont
  {Nisoli}},\ }\href {\doibase 10.1126/science.1254061} {\bibfield  {journal}
  {\bibinfo  {journal} {Science}\ }\textbf {\bibinfo {volume} {346}},\ \bibinfo
  {pages} {336} (\bibinfo {year} {2014})}\BibitemShut {NoStop}%
\bibitem [{\citenamefont {Jager}\ \emph {et~al.}(2017)\citenamefont {Jager},
  \citenamefont {Ott}, \citenamefont {Kraus}, \citenamefont {Kaplan},
  \citenamefont {Pouse}, \citenamefont {Marvel}, \citenamefont {Haglund},
  \citenamefont {Neumark},\ and\ \citenamefont {Leone}}]{Jager2017}%
  \BibitemOpen
  \bibfield  {author} {\bibinfo {author} {\bibfnamefont {M.~F.}\ \bibnamefont
  {Jager}}, \bibinfo {author} {\bibfnamefont {C.}~\bibnamefont {Ott}}, \bibinfo
  {author} {\bibfnamefont {P.~M.}\ \bibnamefont {Kraus}}, \bibinfo {author}
  {\bibfnamefont {C.~J.}\ \bibnamefont {Kaplan}}, \bibinfo {author}
  {\bibfnamefont {W.}~\bibnamefont {Pouse}}, \bibinfo {author} {\bibfnamefont
  {R.~E.}\ \bibnamefont {Marvel}}, \bibinfo {author} {\bibfnamefont {R.~F.}\
  \bibnamefont {Haglund}}, \bibinfo {author} {\bibfnamefont {D.~M.}\
  \bibnamefont {Neumark}}, \ and\ \bibinfo {author} {\bibfnamefont {S.~R.}\
  \bibnamefont {Leone}},\ }\href {\doibase 10.1073/pnas.1707602114} {\bibfield
  {journal} {\bibinfo  {journal} {Proceedings of the National Academy of
  Sciences}\ }\textbf {\bibinfo {volume} {114}},\ \bibinfo {pages} {9558}
  (\bibinfo {year} {2017})}\BibitemShut {NoStop}%
\bibitem [{\citenamefont {Kaplan}\ \emph {et~al.}(2018)\citenamefont {Kaplan},
  \citenamefont {Kraus}, \citenamefont {Ross}, \citenamefont {Z{\"{u}}rch},
  \citenamefont {Cushing}, \citenamefont {Jager}, \citenamefont {Chang},
  \citenamefont {Gullikson}, \citenamefont {Neumark},\ and\ \citenamefont
  {Leone}}]{Kaplan2018}%
  \BibitemOpen
  \bibfield  {author} {\bibinfo {author} {\bibfnamefont {C.~J.}\ \bibnamefont
  {Kaplan}}, \bibinfo {author} {\bibfnamefont {P.~M.}\ \bibnamefont {Kraus}},
  \bibinfo {author} {\bibfnamefont {A.~D.}\ \bibnamefont {Ross}}, \bibinfo
  {author} {\bibfnamefont {M.}~\bibnamefont {Z{\"{u}}rch}}, \bibinfo {author}
  {\bibfnamefont {S.~K.}\ \bibnamefont {Cushing}}, \bibinfo {author}
  {\bibfnamefont {M.~F.}\ \bibnamefont {Jager}}, \bibinfo {author}
  {\bibfnamefont {H.~T.}\ \bibnamefont {Chang}}, \bibinfo {author}
  {\bibfnamefont {E.~M.}\ \bibnamefont {Gullikson}}, \bibinfo {author}
  {\bibfnamefont {D.~M.}\ \bibnamefont {Neumark}}, \ and\ \bibinfo {author}
  {\bibfnamefont {S.~R.}\ \bibnamefont {Leone}},\ }\href {\doibase
  10.1103/PhysRevB.97.205202} {\bibfield  {journal} {\bibinfo  {journal}
  {Physical Review B}\ }\textbf {\bibinfo {volume} {97}},\ \bibinfo {pages}
  {205202} (\bibinfo {year} {2018})}\BibitemShut {NoStop}%
\bibitem [{\citenamefont {Zhang}\ \emph {et~al.}(2016)\citenamefont {Zhang},
  \citenamefont {Lin}, \citenamefont {Ryland}, \citenamefont {Verkamp},
  \citenamefont {Benke}, \citenamefont {{De Groot}}, \citenamefont {Girolami},\
  and\ \citenamefont {Vura-Weis}}]{Zhang2016}%
  \BibitemOpen
  \bibfield  {author} {\bibinfo {author} {\bibfnamefont {K.}~\bibnamefont
  {Zhang}}, \bibinfo {author} {\bibfnamefont {M.~F.}\ \bibnamefont {Lin}},
  \bibinfo {author} {\bibfnamefont {E.~S.}\ \bibnamefont {Ryland}}, \bibinfo
  {author} {\bibfnamefont {M.~A.}\ \bibnamefont {Verkamp}}, \bibinfo {author}
  {\bibfnamefont {K.}~\bibnamefont {Benke}}, \bibinfo {author} {\bibfnamefont
  {F.~M.}\ \bibnamefont {{De Groot}}}, \bibinfo {author} {\bibfnamefont
  {G.~S.}\ \bibnamefont {Girolami}}, \ and\ \bibinfo {author} {\bibfnamefont
  {J.}~\bibnamefont {Vura-Weis}},\ }\href {\doibase
  10.1021/acs.jpclett.6b01393} {\bibfield  {journal} {\bibinfo  {journal}
  {Journal of Physical Chemistry Letters}\ }\textbf {\bibinfo {volume} {7}},\
  \bibinfo {pages} {3383} (\bibinfo {year} {2016})}\BibitemShut {NoStop}%
\bibitem [{\citenamefont {Kraus}\ \emph {et~al.}(2015)\citenamefont {Kraus},
  \citenamefont {Mignolet}, \citenamefont {Baykusheva}, \citenamefont
  {Rupenyan}, \citenamefont {Horn{\'{y}}}, \citenamefont {Penka}, \citenamefont
  {Grassi}, \citenamefont {Tolstikhin}, \citenamefont {Schneider},
  \citenamefont {Jensen}, \citenamefont {Madsen}, \citenamefont {Bandrauk},
  \citenamefont {Remacle},\ and\ \citenamefont {W{\"{o}}rner}}]{Kraus2015}%
  \BibitemOpen
  \bibfield  {author} {\bibinfo {author} {\bibfnamefont {P.~M.}\ \bibnamefont
  {Kraus}}, \bibinfo {author} {\bibfnamefont {B.}~\bibnamefont {Mignolet}},
  \bibinfo {author} {\bibfnamefont {D.}~\bibnamefont {Baykusheva}}, \bibinfo
  {author} {\bibfnamefont {A.}~\bibnamefont {Rupenyan}}, \bibinfo {author}
  {\bibfnamefont {L.}~\bibnamefont {Horn{\'{y}}}}, \bibinfo {author}
  {\bibfnamefont {E.~F.}\ \bibnamefont {Penka}}, \bibinfo {author}
  {\bibfnamefont {G.}~\bibnamefont {Grassi}}, \bibinfo {author} {\bibfnamefont
  {O.~I.}\ \bibnamefont {Tolstikhin}}, \bibinfo {author} {\bibfnamefont
  {J.}~\bibnamefont {Schneider}}, \bibinfo {author} {\bibfnamefont
  {F.}~\bibnamefont {Jensen}}, \bibinfo {author} {\bibfnamefont {L.~B.}\
  \bibnamefont {Madsen}}, \bibinfo {author} {\bibfnamefont {A.~D.}\
  \bibnamefont {Bandrauk}}, \bibinfo {author} {\bibfnamefont {F.}~\bibnamefont
  {Remacle}}, \ and\ \bibinfo {author} {\bibfnamefont {H.~J.}\ \bibnamefont
  {W{\"{o}}rner}},\ }\href {\doibase 10.1126/science.aab2160} {\bibfield
  {journal} {\bibinfo  {journal} {Science}\ }\textbf {\bibinfo {volume}
  {350}},\ \bibinfo {pages} {790} (\bibinfo {year} {2015})}\BibitemShut
  {NoStop}%
\bibitem [{\citenamefont {W{\"{o}}rner}\ \emph {et~al.}(2017)\citenamefont
  {W{\"{o}}rner}, \citenamefont {Arrell}, \citenamefont {Banerji},
  \citenamefont {Cannizzo}, \citenamefont {Chergui}, \citenamefont {Das},
  \citenamefont {Hamm}, \citenamefont {Keller}, \citenamefont {Kraus},
  \citenamefont {Liberatore}, \citenamefont {Lopez-Tarifa}, \citenamefont
  {Lucchini}, \citenamefont {Meuwly}, \citenamefont {Milne}, \citenamefont
  {Moser}, \citenamefont {Rothlisberger}, \citenamefont {Smolentsev},
  \citenamefont {Teuscher}, \citenamefont {{Van Bokhoven}},\ and\ \citenamefont
  {Wenger}}]{Worner2017}%
  \BibitemOpen
  \bibfield  {author} {\bibinfo {author} {\bibfnamefont {H.~J.}\ \bibnamefont
  {W{\"{o}}rner}}, \bibinfo {author} {\bibfnamefont {C.~A.}\ \bibnamefont
  {Arrell}}, \bibinfo {author} {\bibfnamefont {N.}~\bibnamefont {Banerji}},
  \bibinfo {author} {\bibfnamefont {A.}~\bibnamefont {Cannizzo}}, \bibinfo
  {author} {\bibfnamefont {M.}~\bibnamefont {Chergui}}, \bibinfo {author}
  {\bibfnamefont {A.~K.}\ \bibnamefont {Das}}, \bibinfo {author} {\bibfnamefont
  {P.}~\bibnamefont {Hamm}}, \bibinfo {author} {\bibfnamefont {U.}~\bibnamefont
  {Keller}}, \bibinfo {author} {\bibfnamefont {P.~M.}\ \bibnamefont {Kraus}},
  \bibinfo {author} {\bibfnamefont {E.}~\bibnamefont {Liberatore}}, \bibinfo
  {author} {\bibfnamefont {P.}~\bibnamefont {Lopez-Tarifa}}, \bibinfo {author}
  {\bibfnamefont {M.}~\bibnamefont {Lucchini}}, \bibinfo {author}
  {\bibfnamefont {M.}~\bibnamefont {Meuwly}}, \bibinfo {author} {\bibfnamefont
  {C.}~\bibnamefont {Milne}}, \bibinfo {author} {\bibfnamefont {J.~E.}\
  \bibnamefont {Moser}}, \bibinfo {author} {\bibfnamefont {U.}~\bibnamefont
  {Rothlisberger}}, \bibinfo {author} {\bibfnamefont {G.}~\bibnamefont
  {Smolentsev}}, \bibinfo {author} {\bibfnamefont {J.}~\bibnamefont
  {Teuscher}}, \bibinfo {author} {\bibfnamefont {J.~A.}\ \bibnamefont {{Van
  Bokhoven}}}, \ and\ \bibinfo {author} {\bibfnamefont {O.}~\bibnamefont
  {Wenger}},\ }\href {\doibase 10.1063/1.4996505} {\bibfield  {journal}
  {\bibinfo  {journal} {Structural Dynamics}\ }\textbf {\bibinfo {volume} {4}}
  (\bibinfo {year} {2017}),\ 10.1063/1.4996505}\BibitemShut {NoStop}%
\bibitem [{\citenamefont {Ryland}\ \emph {et~al.}(2018)\citenamefont {Ryland},
  \citenamefont {Lin}, \citenamefont {Verkamp}, \citenamefont {Zhang},
  \citenamefont {Benke}, \citenamefont {Carlson},\ and\ \citenamefont
  {Vura-Weis}}]{Ryland2018}%
  \BibitemOpen
  \bibfield  {author} {\bibinfo {author} {\bibfnamefont {E.~S.}\ \bibnamefont
  {Ryland}}, \bibinfo {author} {\bibfnamefont {M.~F.}\ \bibnamefont {Lin}},
  \bibinfo {author} {\bibfnamefont {M.~A.}\ \bibnamefont {Verkamp}}, \bibinfo
  {author} {\bibfnamefont {K.}~\bibnamefont {Zhang}}, \bibinfo {author}
  {\bibfnamefont {K.}~\bibnamefont {Benke}}, \bibinfo {author} {\bibfnamefont
  {M.}~\bibnamefont {Carlson}}, \ and\ \bibinfo {author} {\bibfnamefont
  {J.}~\bibnamefont {Vura-Weis}},\ }\href {\doibase 10.1021/jacs.8b01101}
  {\bibfield  {journal} {\bibinfo  {journal} {Journal of the American Chemical
  Society}\ }\textbf {\bibinfo {volume} {140}},\ \bibinfo {pages} {4691}
  (\bibinfo {year} {2018})}\BibitemShut {NoStop}%
\bibitem [{\citenamefont {Zhang}\ \emph {et~al.}(2019)\citenamefont {Zhang},
  \citenamefont {Ash}, \citenamefont {Girolami},\ and\ \citenamefont
  {Vura-Weis}}]{Zhang2019}%
  \BibitemOpen
  \bibfield  {author} {\bibinfo {author} {\bibfnamefont {K.}~\bibnamefont
  {Zhang}}, \bibinfo {author} {\bibfnamefont {R.}~\bibnamefont {Ash}}, \bibinfo
  {author} {\bibfnamefont {G.~S.}\ \bibnamefont {Girolami}}, \ and\ \bibinfo
  {author} {\bibfnamefont {J.}~\bibnamefont {Vura-Weis}},\ }\href {\doibase
  10.1021/jacs.9b07332} {\bibfield  {journal} {\bibinfo  {journal} {Journal of
  the American Chemical Society}\ }\textbf {\bibinfo {volume} {141}},\ \bibinfo
  {pages} {17180} (\bibinfo {year} {2019})}\BibitemShut {NoStop}%
\bibitem [{\citenamefont {Z{\"{u}}rch}\ \emph {et~al.}(2017)\citenamefont
  {Z{\"{u}}rch}, \citenamefont {Chang}, \citenamefont {Borja}, \citenamefont
  {Kraus}, \citenamefont {Cushing}, \citenamefont {Gandman}, \citenamefont
  {Kaplan}, \citenamefont {Oh}, \citenamefont {Prell}, \citenamefont
  {Prendergast}, \citenamefont {Pemmaraju}, \citenamefont {Neumark},\ and\
  \citenamefont {Leone}}]{Zurch2017a}%
  \BibitemOpen
  \bibfield  {author} {\bibinfo {author} {\bibfnamefont {M.}~\bibnamefont
  {Z{\"{u}}rch}}, \bibinfo {author} {\bibfnamefont {H.~T.}\ \bibnamefont
  {Chang}}, \bibinfo {author} {\bibfnamefont {L.~J.}\ \bibnamefont {Borja}},
  \bibinfo {author} {\bibfnamefont {P.~M.}\ \bibnamefont {Kraus}}, \bibinfo
  {author} {\bibfnamefont {S.~K.}\ \bibnamefont {Cushing}}, \bibinfo {author}
  {\bibfnamefont {A.}~\bibnamefont {Gandman}}, \bibinfo {author} {\bibfnamefont
  {C.~J.}\ \bibnamefont {Kaplan}}, \bibinfo {author} {\bibfnamefont {M.~H.}\
  \bibnamefont {Oh}}, \bibinfo {author} {\bibfnamefont {J.~S.}\ \bibnamefont
  {Prell}}, \bibinfo {author} {\bibfnamefont {D.}~\bibnamefont {Prendergast}},
  \bibinfo {author} {\bibfnamefont {C.~D.}\ \bibnamefont {Pemmaraju}}, \bibinfo
  {author} {\bibfnamefont {D.~M.}\ \bibnamefont {Neumark}}, \ and\ \bibinfo
  {author} {\bibfnamefont {S.~R.}\ \bibnamefont {Leone}},\ }\href {\doibase
  10.1038/ncomms15734} {\bibfield  {journal} {\bibinfo  {journal} {Nature
  Communications}\ }\textbf {\bibinfo {volume} {8}},\ \bibinfo {pages} {15734}
  (\bibinfo {year} {2017})}\BibitemShut {NoStop}%
\bibitem [{\citenamefont {Cushing}\ \emph {et~al.}(2018)\citenamefont
  {Cushing}, \citenamefont {Z{\"{u}}rch}, \citenamefont {Kraus}, \citenamefont
  {Carneiro}, \citenamefont {Lee}, \citenamefont {Chang}, \citenamefont
  {Kaplan},\ and\ \citenamefont {Leone}}]{Cushing2017}%
  \BibitemOpen
  \bibfield  {author} {\bibinfo {author} {\bibfnamefont {S.~K.}\ \bibnamefont
  {Cushing}}, \bibinfo {author} {\bibfnamefont {M.}~\bibnamefont
  {Z{\"{u}}rch}}, \bibinfo {author} {\bibfnamefont {P.~M.}\ \bibnamefont
  {Kraus}}, \bibinfo {author} {\bibfnamefont {L.~M.}\ \bibnamefont {Carneiro}},
  \bibinfo {author} {\bibfnamefont {A.}~\bibnamefont {Lee}}, \bibinfo {author}
  {\bibfnamefont {H.~T.}\ \bibnamefont {Chang}}, \bibinfo {author}
  {\bibfnamefont {C.~J.}\ \bibnamefont {Kaplan}}, \ and\ \bibinfo {author}
  {\bibfnamefont {S.~R.}\ \bibnamefont {Leone}},\ }\href {\doibase
  10.1063/1.5038015} {\bibfield  {journal} {\bibinfo  {journal} {Structural
  Dynamics}\ }\textbf {\bibinfo {volume} {5}},\ \bibinfo {pages} {1} (\bibinfo
  {year} {2018})}\BibitemShut {NoStop}%
\bibitem [{\citenamefont {Cushing}\ \emph {et~al.}(2020)\citenamefont
  {Cushing}, \citenamefont {Porter}, \citenamefont {de~Roulet}, \citenamefont
  {Lee}, \citenamefont {Marsh}, \citenamefont {Szoke}, \citenamefont {Vaida},\
  and\ \citenamefont {Leone}}]{Cushing2020}%
  \BibitemOpen
  \bibfield  {author} {\bibinfo {author} {\bibfnamefont {S.~K.}\ \bibnamefont
  {Cushing}}, \bibinfo {author} {\bibfnamefont {I.~J.}\ \bibnamefont {Porter}},
  \bibinfo {author} {\bibfnamefont {B.~R.}\ \bibnamefont {de~Roulet}}, \bibinfo
  {author} {\bibfnamefont {A.}~\bibnamefont {Lee}}, \bibinfo {author}
  {\bibfnamefont {B.~M.}\ \bibnamefont {Marsh}}, \bibinfo {author}
  {\bibfnamefont {S.}~\bibnamefont {Szoke}}, \bibinfo {author} {\bibfnamefont
  {M.~E.}\ \bibnamefont {Vaida}}, \ and\ \bibinfo {author} {\bibfnamefont
  {S.~R.}\ \bibnamefont {Leone}},\ }\href {\doibase 10.1126/sciadv.aay6650}
  {\bibfield  {journal} {\bibinfo  {journal} {Science Advances}\ }\textbf
  {\bibinfo {volume} {6}} (\bibinfo {year} {2020}),\
  10.1126/sciadv.aay6650}\BibitemShut {NoStop}%
\bibitem [{\citenamefont {Li}\ \emph {et~al.}(2017)\citenamefont {Li},
  \citenamefont {Liu}, \citenamefont {Pal}, \citenamefont {Wang}, \citenamefont
  {Ober},\ and\ \citenamefont {Giannelis}}]{Li2017}%
  \BibitemOpen
  \bibfield  {author} {\bibinfo {author} {\bibfnamefont {L.}~\bibnamefont
  {Li}}, \bibinfo {author} {\bibfnamefont {X.}~\bibnamefont {Liu}}, \bibinfo
  {author} {\bibfnamefont {S.}~\bibnamefont {Pal}}, \bibinfo {author}
  {\bibfnamefont {S.}~\bibnamefont {Wang}}, \bibinfo {author} {\bibfnamefont
  {C.~K.}\ \bibnamefont {Ober}}, \ and\ \bibinfo {author} {\bibfnamefont
  {E.~P.}\ \bibnamefont {Giannelis}},\ }\href {\doibase 10.1039/c7cs00080d}
  {\bibfield  {journal} {\bibinfo  {journal} {Chemical Society Reviews}\
  }\textbf {\bibinfo {volume} {46}},\ \bibinfo {pages} {4855} (\bibinfo {year}
  {2017})}\BibitemShut {NoStop}%
\bibitem [{\citenamefont {Haitjema}\ \emph {et~al.}(2017)\citenamefont
  {Haitjema}, \citenamefont {Zhang}, \citenamefont {Vockenhuber}, \citenamefont
  {Kazazis}, \citenamefont {Ekinci},\ and\ \citenamefont
  {Brouwer}}]{Haitjema2017a}%
  \BibitemOpen
  \bibfield  {author} {\bibinfo {author} {\bibfnamefont {J.}~\bibnamefont
  {Haitjema}}, \bibinfo {author} {\bibfnamefont {Y.}~\bibnamefont {Zhang}},
  \bibinfo {author} {\bibfnamefont {M.}~\bibnamefont {Vockenhuber}}, \bibinfo
  {author} {\bibfnamefont {D.}~\bibnamefont {Kazazis}}, \bibinfo {author}
  {\bibfnamefont {Y.}~\bibnamefont {Ekinci}}, \ and\ \bibinfo {author}
  {\bibfnamefont {A.~M.}\ \bibnamefont {Brouwer}},\ }\href {\doibase
  10.1117/1.jmm.16.3.033510} {\bibfield  {journal} {\bibinfo  {journal}
  {Journal of Micro/Nanolithography, MEMS, and MOEMS}\ }\textbf {\bibinfo
  {volume} {16}},\ \bibinfo {pages} {1} (\bibinfo {year} {2017})}\BibitemShut
  {NoStop}%
\bibitem [{\citenamefont {Wu}\ \emph {et~al.}(2019{\natexlab{a}})\citenamefont
  {Wu}, \citenamefont {Liu}, \citenamefont {Vockenhuber}, \citenamefont
  {Ekinci},\ and\ \citenamefont {Castellanos}}]{Wu2019a}%
  \BibitemOpen
  \bibfield  {author} {\bibinfo {author} {\bibfnamefont {L.}~\bibnamefont
  {Wu}}, \bibinfo {author} {\bibfnamefont {J.}~\bibnamefont {Liu}}, \bibinfo
  {author} {\bibfnamefont {M.}~\bibnamefont {Vockenhuber}}, \bibinfo {author}
  {\bibfnamefont {Y.}~\bibnamefont {Ekinci}}, \ and\ \bibinfo {author}
  {\bibfnamefont {S.}~\bibnamefont {Castellanos}},\ }\href {\doibase
  10.1002/ejic.201900745} {\bibfield  {journal} {\bibinfo  {journal} {European
  Journal of Inorganic Chemistry}\ }\textbf {\bibinfo {volume} {2019}},\
  \bibinfo {pages} {4136} (\bibinfo {year} {2019}{\natexlab{a}})}\BibitemShut
  {NoStop}%
\bibitem [{\citenamefont {Sadegh}\ \emph {et~al.}(2020)\citenamefont {Sadegh},
  \citenamefont {van~der Geest}, \citenamefont {Haitjema}, \citenamefont
  {Campi}, \citenamefont {Castellanos}, \citenamefont {Kraus},\ and\
  \citenamefont {Brouwer}}]{Sadegh2020}%
  \BibitemOpen
  \bibfield  {author} {\bibinfo {author} {\bibfnamefont {N.}~\bibnamefont
  {Sadegh}}, \bibinfo {author} {\bibfnamefont {M.}~\bibnamefont {van~der
  Geest}}, \bibinfo {author} {\bibfnamefont {J.}~\bibnamefont {Haitjema}},
  \bibinfo {author} {\bibfnamefont {F.}~\bibnamefont {Campi}}, \bibinfo
  {author} {\bibfnamefont {S.}~\bibnamefont {Castellanos}}, \bibinfo {author}
  {\bibfnamefont {P.~M.}\ \bibnamefont {Kraus}}, \ and\ \bibinfo {author}
  {\bibfnamefont {A.~M.}\ \bibnamefont {Brouwer}},\ }\href {\doibase
  10.2494/photopolymer.33.145} {\bibfield  {journal} {\bibinfo  {journal}
  {Journal of Photopolymer Science and Technology}\ }\textbf {\bibinfo {volume}
  {33}},\ \bibinfo {pages} {145} (\bibinfo {year} {2020})}\BibitemShut
  {NoStop}%
\bibitem [{\citenamefont {Timmers}\ \emph {et~al.}(2014)\citenamefont
  {Timmers}, \citenamefont {Li}, \citenamefont {Shivaram}, \citenamefont
  {Santra}, \citenamefont {Vendrell},\ and\ \citenamefont
  {Sandhu}}]{Timmers2014}%
  \BibitemOpen
  \bibfield  {author} {\bibinfo {author} {\bibfnamefont {H.}~\bibnamefont
  {Timmers}}, \bibinfo {author} {\bibfnamefont {Z.}~\bibnamefont {Li}},
  \bibinfo {author} {\bibfnamefont {N.}~\bibnamefont {Shivaram}}, \bibinfo
  {author} {\bibfnamefont {R.}~\bibnamefont {Santra}}, \bibinfo {author}
  {\bibfnamefont {O.}~\bibnamefont {Vendrell}}, \ and\ \bibinfo {author}
  {\bibfnamefont {A.}~\bibnamefont {Sandhu}},\ }\href {\doibase
  10.1103/PhysRevLett.113.113003} {\bibfield  {journal} {\bibinfo  {journal}
  {Physical Review Letters}\ }\textbf {\bibinfo {volume} {113}},\ \bibinfo
  {pages} {1} (\bibinfo {year} {2014})}\BibitemShut {NoStop}%
\bibitem [{\citenamefont {Marciniak}\ \emph {et~al.}(2015)\citenamefont
  {Marciniak}, \citenamefont {Despr{\'{e}}}, \citenamefont {Barillot},
  \citenamefont {Rouz{\'{e}}e}, \citenamefont {Galbraith}, \citenamefont
  {Klei}, \citenamefont {Yang}, \citenamefont {Smeenk}, \citenamefont {Loriot},
  \citenamefont {Reddy}, \citenamefont {Tielens}, \citenamefont {Mahapatra},
  \citenamefont {Kuleff}, \citenamefont {Vrakking},\ and\ \citenamefont
  {L{\'{e}}pine}}]{Marciniak2015}%
  \BibitemOpen
  \bibfield  {author} {\bibinfo {author} {\bibfnamefont {A.}~\bibnamefont
  {Marciniak}}, \bibinfo {author} {\bibfnamefont {V.}~\bibnamefont
  {Despr{\'{e}}}}, \bibinfo {author} {\bibfnamefont {T.}~\bibnamefont
  {Barillot}}, \bibinfo {author} {\bibfnamefont {A.}~\bibnamefont
  {Rouz{\'{e}}e}}, \bibinfo {author} {\bibfnamefont {M.~C.}\ \bibnamefont
  {Galbraith}}, \bibinfo {author} {\bibfnamefont {J.}~\bibnamefont {Klei}},
  \bibinfo {author} {\bibfnamefont {C.~H.}\ \bibnamefont {Yang}}, \bibinfo
  {author} {\bibfnamefont {C.~T.}\ \bibnamefont {Smeenk}}, \bibinfo {author}
  {\bibfnamefont {V.}~\bibnamefont {Loriot}}, \bibinfo {author} {\bibfnamefont
  {S.~N.}\ \bibnamefont {Reddy}}, \bibinfo {author} {\bibfnamefont {A.~G.}\
  \bibnamefont {Tielens}}, \bibinfo {author} {\bibfnamefont {S.}~\bibnamefont
  {Mahapatra}}, \bibinfo {author} {\bibfnamefont {A.~I.}\ \bibnamefont
  {Kuleff}}, \bibinfo {author} {\bibfnamefont {M.~J.}\ \bibnamefont
  {Vrakking}}, \ and\ \bibinfo {author} {\bibfnamefont {F.}~\bibnamefont
  {L{\'{e}}pine}},\ }\href {\doibase 10.1038/ncomms8909} {\bibfield  {journal}
  {\bibinfo  {journal} {Nature Communications}\ }\textbf {\bibinfo {volume}
  {6}} (\bibinfo {year} {2015}),\ 10.1038/ncomms8909}\BibitemShut {NoStop}%
\bibitem [{\citenamefont {Vielhauer}\ \emph {et~al.}(2009)\citenamefont
  {Vielhauer}, \citenamefont {Babin}, \citenamefont {{De Grazia}},
  \citenamefont {Feldbach}, \citenamefont {Kirm}, \citenamefont {Nagirnyi},\
  and\ \citenamefont {Vasil'ev}}]{Vielhauer2009}%
  \BibitemOpen
  \bibfield  {author} {\bibinfo {author} {\bibfnamefont {S.}~\bibnamefont
  {Vielhauer}}, \bibinfo {author} {\bibfnamefont {V.}~\bibnamefont {Babin}},
  \bibinfo {author} {\bibfnamefont {M.}~\bibnamefont {{De Grazia}}}, \bibinfo
  {author} {\bibfnamefont {E.}~\bibnamefont {Feldbach}}, \bibinfo {author}
  {\bibfnamefont {M.}~\bibnamefont {Kirm}}, \bibinfo {author} {\bibfnamefont
  {V.}~\bibnamefont {Nagirnyi}}, \ and\ \bibinfo {author} {\bibfnamefont
  {A.~N.}\ \bibnamefont {Vasil'ev}},\ }\href {\doibase 10.1117/12.822292}
  {\bibfield  {journal} {\bibinfo  {journal} {Damage to VUV, EUV, and X-Ray
  Optics II}\ }\textbf {\bibinfo {volume} {7361}},\ \bibinfo {pages} {73610R}
  (\bibinfo {year} {2009})}\BibitemShut {NoStop}%
\bibitem [{\citenamefont {Savchyn}\ \emph {et~al.}(2012)\citenamefont
  {Savchyn}, \citenamefont {Vistovskyy}, \citenamefont {Pushak}, \citenamefont
  {Voloshinovskii}, \citenamefont {Gektin}, \citenamefont {Pankratov},\ and\
  \citenamefont {Popov}}]{Savchyn2012}%
  \BibitemOpen
  \bibfield  {author} {\bibinfo {author} {\bibfnamefont {P.~V.}\ \bibnamefont
  {Savchyn}}, \bibinfo {author} {\bibfnamefont {V.~V.}\ \bibnamefont
  {Vistovskyy}}, \bibinfo {author} {\bibfnamefont {A.~S.}\ \bibnamefont
  {Pushak}}, \bibinfo {author} {\bibfnamefont {A.~S.}\ \bibnamefont
  {Voloshinovskii}}, \bibinfo {author} {\bibfnamefont {A.~V.}\ \bibnamefont
  {Gektin}}, \bibinfo {author} {\bibfnamefont {V.}~\bibnamefont {Pankratov}}, \
  and\ \bibinfo {author} {\bibfnamefont {A.~I.}\ \bibnamefont {Popov}},\ }\href
  {\doibase 10.1016/j.nimb.2011.11.024} {\bibfield  {journal} {\bibinfo
  {journal} {Nuclear Instruments and Methods in Physics Research, Section B}}\
  }\textbf {\bibinfo {volume} {274}},\ \bibinfo {pages} {78} (\bibinfo {year}
  {2012})\BibitemShut {NoStop}%
\bibitem [{\citenamefont {Sokolov}\ \emph {et~al.}(2012)\citenamefont
  {Sokolov}, \citenamefont {Pustovarov}, \citenamefont {Churmanov},
  \citenamefont {Ivanov}, \citenamefont {Gruzdev}, \citenamefont {Sokolov},
  \citenamefont {Baranov},\ and\ \citenamefont {Moskvin}}]{Sokolov2012}%
  \BibitemOpen
  \bibfield  {author} {\bibinfo {author} {\bibfnamefont {V.~I.}\ \bibnamefont
  {Sokolov}}, \bibinfo {author} {\bibfnamefont {V.~A.}\ \bibnamefont
  {Pustovarov}}, \bibinfo {author} {\bibfnamefont {V.~N.}\ \bibnamefont
  {Churmanov}}, \bibinfo {author} {\bibfnamefont {V.~Y.}\ \bibnamefont
  {Ivanov}}, \bibinfo {author} {\bibfnamefont {N.~B.}\ \bibnamefont {Gruzdev}},
  \bibinfo {author} {\bibfnamefont {P.~S.}\ \bibnamefont {Sokolov}}, \bibinfo
  {author} {\bibfnamefont {A.~N.}\ \bibnamefont {Baranov}}, \ and\ \bibinfo
  {author} {\bibfnamefont {A.~S.}\ \bibnamefont {Moskvin}},\ }\href {\doibase
  10.1103/PhysRevB.86.115128} {\bibfield  {journal} {\bibinfo  {journal}
  {Physical Review B}\ }\textbf {\bibinfo {volume} {86}},\ \bibinfo {pages}
  {115128} (\bibinfo {year} {2012})}\BibitemShut {NoStop}%
\bibitem [{\citenamefont {Pustovarov}\ \emph {et~al.}(2013)\citenamefont
  {Pustovarov}, \citenamefont {Razumov}, \citenamefont {Ivanov}, \citenamefont
  {Vyprintsev},\ and\ \citenamefont {Shvalev}}]{Pustovarov2013}%
  \BibitemOpen
  \bibfield  {author} {\bibinfo {author} {\bibfnamefont {V.~A.}\ \bibnamefont
  {Pustovarov}}, \bibinfo {author} {\bibfnamefont {A.~N.}\ \bibnamefont
  {Razumov}}, \bibinfo {author} {\bibfnamefont {V.~Y.}\ \bibnamefont {Ivanov}},
  \bibinfo {author} {\bibfnamefont {D.~I.}\ \bibnamefont {Vyprintsev}}, \ and\
  \bibinfo {author} {\bibfnamefont {N.~G.}\ \bibnamefont {Shvalev}},\ }\href
  {\doibase 10.3103/S1062873813020287} {\bibfield  {journal} {\bibinfo
  {journal} {Bulletin of the Russian Academy of Sciences: Physics}\ }\textbf
  {\bibinfo {volume} {77}},\ \bibinfo {pages} {217} (\bibinfo {year}
  {2013})}\BibitemShut {NoStop}%
\bibitem [{\citenamefont {Bel'skii}\ \emph {et~al.}(2016)\citenamefont
  {Bel'skii}, \citenamefont {Vasil'ev}, \citenamefont {Ivanov}, \citenamefont
  {Kamenskikh}, \citenamefont {Kolobanov}, \citenamefont {Makhov},\ and\
  \citenamefont {Spasskii}}]{Belskii2016}%
  \BibitemOpen
  \bibfield  {author} {\bibinfo {author} {\bibfnamefont {A.~N.}\ \bibnamefont
  {Bel'skii}}, \bibinfo {author} {\bibfnamefont {A.~N.}\ \bibnamefont
  {Vasil'ev}}, \bibinfo {author} {\bibfnamefont {S.~N.}\ \bibnamefont
  {Ivanov}}, \bibinfo {author} {\bibfnamefont {I.~A.}\ \bibnamefont
  {Kamenskikh}}, \bibinfo {author} {\bibfnamefont {V.~N.}\ \bibnamefont
  {Kolobanov}}, \bibinfo {author} {\bibfnamefont {V.~N.}\ \bibnamefont
  {Makhov}}, \ and\ \bibinfo {author} {\bibfnamefont {D.~A.}\ \bibnamefont
  {Spasskii}},\ }\href {\doibase 10.1134/S1063774516060043} {\bibfield
  {journal} {\bibinfo  {journal} {Crystallography Reports}\ }\textbf {\bibinfo
  {volume} {61}},\ \bibinfo {pages} {886} (\bibinfo {year} {2016})}\BibitemShut
  {NoStop}%
\bibitem [{\citenamefont {Krzywinski}\ \emph {et~al.}(2017)\citenamefont
  {Krzywinski}, \citenamefont {Andrejczuk}, \citenamefont {Bionta},
  \citenamefont {Burian}, \citenamefont {Chalupsk{\'{y}}}, \citenamefont
  {Jurek}, \citenamefont {Kirm}, \citenamefont {Nagirnyi}, \citenamefont
  {Sobierajski}, \citenamefont {Tiedtke}, \citenamefont {Vielhauer},\ and\
  \citenamefont {Juha}}]{Krzywinski2017}%
  \BibitemOpen
  \bibfield  {author} {\bibinfo {author} {\bibfnamefont {J.}~\bibnamefont
  {Krzywinski}}, \bibinfo {author} {\bibfnamefont {A.}~\bibnamefont
  {Andrejczuk}}, \bibinfo {author} {\bibfnamefont {R.~M.}\ \bibnamefont
  {Bionta}}, \bibinfo {author} {\bibfnamefont {T.}~\bibnamefont {Burian}},
  \bibinfo {author} {\bibfnamefont {J.}~\bibnamefont {Chalupsk{\'{y}}}},
  \bibinfo {author} {\bibfnamefont {M.}~\bibnamefont {Jurek}}, \bibinfo
  {author} {\bibfnamefont {M.}~\bibnamefont {Kirm}}, \bibinfo {author}
  {\bibfnamefont {V.}~\bibnamefont {Nagirnyi}}, \bibinfo {author}
  {\bibfnamefont {R.}~\bibnamefont {Sobierajski}}, \bibinfo {author}
  {\bibfnamefont {K.}~\bibnamefont {Tiedtke}}, \bibinfo {author} {\bibfnamefont
  {S.}~\bibnamefont {Vielhauer}}, \ and\ \bibinfo {author} {\bibfnamefont
  {L.}~\bibnamefont {Juha}},\ }\href {\doibase 10.1364/ome.7.000665} {\bibfield
   {journal} {\bibinfo  {journal} {Optical Materials Express}\ }\textbf
  {\bibinfo {volume} {7}},\ \bibinfo {pages} {665} (\bibinfo {year}
  {2017})}\BibitemShut {NoStop}%
\bibitem [{\citenamefont {Chylii}\ \emph {et~al.}(2019)\citenamefont {Chylii},
  \citenamefont {Malyi}, \citenamefont {Rovetskyi}, \citenamefont {Demkiv},
  \citenamefont {Vistovskyy}, \citenamefont {Rodnyi}, \citenamefont {Gektin},
  \citenamefont {Vasil'ev},\ and\ \citenamefont {Voloshinovskii}}]{Chylii2019}%
  \BibitemOpen
  \bibfield  {author} {\bibinfo {author} {\bibfnamefont {M.}~\bibnamefont
  {Chylii}}, \bibinfo {author} {\bibfnamefont {T.}~\bibnamefont {Malyi}},
  \bibinfo {author} {\bibfnamefont {I.}~\bibnamefont {Rovetskyi}}, \bibinfo
  {author} {\bibfnamefont {T.}~\bibnamefont {Demkiv}}, \bibinfo {author}
  {\bibfnamefont {V.}~\bibnamefont {Vistovskyy}}, \bibinfo {author}
  {\bibfnamefont {P.}~\bibnamefont {Rodnyi}}, \bibinfo {author} {\bibfnamefont
  {A.}~\bibnamefont {Gektin}}, \bibinfo {author} {\bibfnamefont
  {A.}~\bibnamefont {Vasil'ev}}, \ and\ \bibinfo {author} {\bibfnamefont
  {A.}~\bibnamefont {Voloshinovskii}},\ }\href {\doibase
  10.1016/j.optmat.2019.03.011} {\bibfield  {journal} {\bibinfo  {journal}
  {Optical Materials}\ }\textbf {\bibinfo {volume} {91}},\ \bibinfo {pages}
  {115} (\bibinfo {year} {2019})}\BibitemShut {NoStop}%
\bibitem [{\citenamefont {Pankratov}\ and\ \citenamefont
  {Kotlov}(2020)}]{Pankratov2020}%
  \BibitemOpen
  \bibfield  {author} {\bibinfo {author} {\bibfnamefont {V.}~\bibnamefont
  {Pankratov}}\ and\ \bibinfo {author} {\bibfnamefont {A.}~\bibnamefont
  {Kotlov}},\ }\href {\doibase 10.1016/j.nimb.2020.04.015}\\ {\bibfield
  {journal} {\bibinfo  {journal} {Nuclear Instruments and Methods in Physics
  Research, Section B}\ }}\textbf {\bibinfo {volume} {474}},\ \bibinfo {pages}
  {35} (\bibinfo {year} {2020})\BibitemShut {NoStop}%
\bibitem [{\citenamefont {Jaworowski}\ \emph {et~al.}(1968)\citenamefont
  {Jaworowski}, \citenamefont {Cosgrove}, \citenamefont {Bracco},\ and\
  \citenamefont {Walters}}]{Jaworowski1968}%
  \BibitemOpen
  \bibfield  {author} {\bibinfo {author} {\bibfnamefont {R.~J.}\ \bibnamefont
  {Jaworowski}}, \bibinfo {author} {\bibfnamefont {J.~F.}\ \bibnamefont
  {Cosgrove}}, \bibinfo {author} {\bibfnamefont {D.~J.}\ \bibnamefont
  {Bracco}}, \ and\ \bibinfo {author} {\bibfnamefont {R.~M.}\ \bibnamefont
  {Walters}},\ }\href {\doibase 10.1016/0584-8547(68)80054-0} {\bibfield
  {journal} {\bibinfo  {journal} {Spectrochimica Acta Part B}\ }\textbf {\bibinfo {volume} {23}},\ \bibinfo {pages} {751}
  (\bibinfo {year} {1968})}\BibitemShut {NoStop}%
\bibitem [{\citenamefont {Soderholm}\ \emph {et~al.}(1998)\citenamefont
  {Soderholm}, \citenamefont {Liu}, \citenamefont {Antonio},\ and\
  \citenamefont {Lytle}}]{Soderholm1998}%
  \BibitemOpen
  \bibfield  {author} {\bibinfo {author} {\bibfnamefont {L.}~\bibnamefont
  {Soderholm}}, \bibinfo {author} {\bibfnamefont {G.~K.}\ \bibnamefont {Liu}},
  \bibinfo {author} {\bibfnamefont {M.~R.}\ \bibnamefont {Antonio}}, \ and\
  \bibinfo {author} {\bibfnamefont {F.~W.}\ \bibnamefont {Lytle}},\ }\href
  {\doibase 10.1063/1.477320} {\bibfield  {journal} {\bibinfo  {journal}
  {Journal of Chemical Physics}\ }\textbf {\bibinfo {volume} {109}},\ \bibinfo
  {pages} {6745} (\bibinfo {year} {1998})}\BibitemShut {NoStop}%
\bibitem [{\citenamefont {Rogalev}\ and\ \citenamefont
  {Goulon}(2002)}]{Rogalev2002}%
  \BibitemOpen
  \bibfield  {author} {\bibinfo {author} {\bibfnamefont {A.}~\bibnamefont
  {Rogalev}}\ and\ \bibinfo {author} {\bibfnamefont {J.}~\bibnamefont
  {Goulon}},\ }in\\ \href {\doibase 10.1142/9789812775757_0015} {\emph {\bibinfo
  {booktitle} {Chemical Applications of Synchrotron Radiation P.II}}}\
  (\bibinfo  {publisher} {World Scientific},\ \bibinfo {year} {2002})\ pp.\
  \bibinfo {pages} {707--760}\BibitemShut {NoStop}%
\bibitem [{\citenamefont {O'Malley}\ \emph {et~al.}(2011)\citenamefont
  {O'Malley}, \citenamefont {Revesz}, \citenamefont {Kazimirov},\ and\
  \citenamefont {Sirenko}}]{OMalley2011}%
  \BibitemOpen
  \bibfield  {author} {\bibinfo {author} {\bibfnamefont {S.~M.}\ \bibnamefont
  {O'Malley}}, \bibinfo {author} {\bibfnamefont {P.}~\bibnamefont {Revesz}},
  \bibinfo {author} {\bibfnamefont {A.}~\bibnamefont {Kazimirov}}, \ and\
  \bibinfo {author} {\bibfnamefont {A.~A.}\ \bibnamefont {Sirenko}},\ }\href
  {\doibase 10.1063/1.3598137} {\bibfield  {journal} {\bibinfo  {journal}
  {Journal of Applied Physics}\ }\textbf {\bibinfo {volume} {109}},\ \bibinfo
  {pages} {124906} (\bibinfo {year} {2011})}\BibitemShut {NoStop}%
\bibitem [{\citenamefont {Pankratov}\ \emph {et~al.}(2011)\citenamefont
  {Pankratov}, \citenamefont {Osinniy}, \citenamefont {Kotlov}, \citenamefont
  {{Nylandsted Larsen}},\ and\ \citenamefont {{Bech Nielsen}}}]{Pankratov2011}%
  \BibitemOpen
  \bibfield  {author} {\bibinfo {author} {\bibfnamefont {V.}~\bibnamefont
  {Pankratov}}, \bibinfo {author} {\bibfnamefont {V.}~\bibnamefont {Osinniy}},
  \bibinfo {author} {\bibfnamefont {A.}~\bibnamefont {Kotlov}}, \bibinfo
  {author} {\bibfnamefont {A.}~\bibnamefont {{Nylandsted Larsen}}}, \ and\
  \bibinfo {author} {\bibfnamefont {B.}~\bibnamefont {{Bech Nielsen}}},\ }\href
  {\doibase 10.1103/PhysRevB.83.045308} {\bibfield  {journal} {\bibinfo
  {journal} {Physical Review B}\ }\textbf {\bibinfo {volume} {83}},\ \bibinfo
  {pages} {1} (\bibinfo {year} {2011})}\BibitemShut {NoStop}%
\bibitem [{\citenamefont {Sekikawa}\ \emph {et~al.}(2000)\citenamefont
  {Sekikawa}, \citenamefont {Ohno}, \citenamefont {Nabekawa},\ and\
  \citenamefont {Watanabe}}]{Sekikawa2000}%
  \BibitemOpen
  \bibfield  {author} {\bibinfo {author} {\bibfnamefont {T.}~\bibnamefont
  {Sekikawa}}, \bibinfo {author} {\bibfnamefont {T.}~\bibnamefont {Ohno}},
  \bibinfo {author} {\bibfnamefont {Y.}~\bibnamefont {Nabekawa}}, \ and\
  \bibinfo {author} {\bibfnamefont {S.}~\bibnamefont {Watanabe}},\ }\href
  {\doibase 10.1016/S0022-2313(99)00429-9} {\bibfield  {journal} {\bibinfo
  {journal} {Journal of Luminescence}\ }\textbf {\bibinfo {volume} {87}},\
  \bibinfo {pages} {827} (\bibinfo {year} {2000})}\BibitemShut {NoStop}%
\bibitem [{\citenamefont {Sekikawa}\ \emph {et~al.}(2002)\citenamefont
  {Sekikawa}, \citenamefont {Yamazaki}, \citenamefont {Nabekawa},\ and\
  \citenamefont {Watanabe}}]{Sekikawa2002}%
  \BibitemOpen
  \bibfield  {author} {\bibinfo {author} {\bibfnamefont {T.}~\bibnamefont
  {Sekikawa}}, \bibinfo {author} {\bibfnamefont {T.}~\bibnamefont {Yamazaki}},
  \bibinfo {author} {\bibfnamefont {Y.}~\bibnamefont {Nabekawa}}, \ and\
  \bibinfo {author} {\bibfnamefont {S.}~\bibnamefont {Watanabe}},\ }\href
  {\doibase 10.1364/josab.19.001941} {\bibfield  {journal} {\bibinfo  {journal}
  {Journal of the Optical Society of America B}\ }\textbf {\bibinfo {volume}
  {19}},\ \bibinfo {pages} {1941} (\bibinfo {year} {2002})}\BibitemShut
  {NoStop}%
\bibitem [{\citenamefont {Kirm}\ \emph {et~al.}(2007)\citenamefont {Kirm},
  \citenamefont {Babin}, \citenamefont {Feldbach}, \citenamefont {Nagirnyi},
  \citenamefont {Vielhauer}, \citenamefont {Guizard}, \citenamefont
  {Carr{\'{e}}}, \citenamefont {{De Grazia}}, \citenamefont {Merdji},
  \citenamefont {Belsky}, \citenamefont {Fedorov},\ and\ \citenamefont
  {Martin}}]{Kirm2007}%
  \BibitemOpen
  \bibfield  {author} {\bibinfo {author} {\bibfnamefont {M.}~\bibnamefont
  {Kirm}}, \bibinfo {author} {\bibfnamefont {V.}~\bibnamefont {Babin}},
  \bibinfo {author} {\bibfnamefont {E.}~\bibnamefont {Feldbach}}, \bibinfo
  {author} {\bibfnamefont {V.}~\bibnamefont {Nagirnyi}}, \bibinfo {author}
  {\bibfnamefont {S.}~\bibnamefont {Vielhauer}}, \bibinfo {author}
  {\bibfnamefont {S.}~\bibnamefont {Guizard}}, \bibinfo {author} {\bibfnamefont
  {B.}~\bibnamefont {Carr{\'{e}}}}, \bibinfo {author} {\bibfnamefont {M.~D.}\
  \bibnamefont {{De Grazia}}}, \bibinfo {author} {\bibfnamefont
  {H.}~\bibnamefont {Merdji}}, \bibinfo {author} {\bibfnamefont {A.~N.}\
  \bibnamefont {Belsky}}, \bibinfo {author} {\bibfnamefont {N.}~\bibnamefont
  {Fedorov}}, \ and\ \bibinfo {author} {\bibfnamefont {P.}~\bibnamefont
  {Martin}},\ }in\ \href {\doibase 10.1002/pssc.200673876} {\emph {\bibinfo
  {booktitle} {Physica Status Solidi (C) Current Topics in Solid State
  Physics}}},\ Vol.~\bibinfo {volume} {4}\ (\bibinfo {year} {2007})\ pp.\
  \bibinfo {pages} {870--876}\BibitemShut {NoStop}%
\bibitem [{\citenamefont {Kickelbick}\ and\ \citenamefont
  {Schubert}(1997)}]{Kickelbick1997}%
  \BibitemOpen
  \bibfield  {author} {\bibinfo {author} {\bibfnamefont {G.}~\bibnamefont
  {Kickelbick}}\ and\ \bibinfo {author} {\bibfnamefont {U.}~\bibnamefont
  {Schubert}},\ }\href {\doibase 10.1002/cber.19971300406} {\bibfield
  {journal} {\bibinfo  {journal} {Chemische Berichte}\ }\textbf {\bibinfo
  {volume} {130}},\ \bibinfo {pages} {473} (\bibinfo {year}
  {1997})}\BibitemShut {NoStop}%
\bibitem [{\citenamefont {Wu}\ \emph {et~al.}(2019{\natexlab{b}})\citenamefont
  {Wu}, \citenamefont {Vockenhuber}, \citenamefont {Ekinci},\ and\
  \citenamefont {{Castellanos Ortega}}}]{Wu2019}%
  \BibitemOpen
  \bibfield  {author} {\bibinfo {author} {\bibfnamefont {L.}~\bibnamefont
  {Wu}}, \bibinfo {author} {\bibfnamefont {M.}~\bibnamefont {Vockenhuber}},
  \bibinfo {author} {\bibfnamefont {Y.}~\bibnamefont {Ekinci}}, \ and\ \bibinfo
  {author} {\bibfnamefont {S.}~\bibnamefont {{Castellanos Ortega}}},\ }in\
  \href {\doibase 10.1117/12.2515264} {\emph {\bibinfo {booktitle} {Proceedings
  of SPIE}}},\ \bibinfo {series and number} {\bibinfo {number} {March}}\
  (\bibinfo {year} {2019})\ p.~\bibinfo {pages} {7}\BibitemShut {NoStop}%
\bibitem [{\citenamefont {Kim}\ \emph {et~al.}(2005)\citenamefont {Kim},
  \citenamefont {Kim}, \citenamefont {Kim}, \citenamefont {Lee}, \citenamefont
  {Lee}, \citenamefont {Park}, \citenamefont {Cho},\ and\ \citenamefont
  {Nam}}]{Kim2005}%
  \BibitemOpen
  \bibfield  {author} {\bibinfo {author} {\bibfnamefont {I.~J.}\ \bibnamefont
  {Kim}}, \bibinfo {author} {\bibfnamefont {C.~M.}\ \bibnamefont {Kim}},
  \bibinfo {author} {\bibfnamefont {H.~T.}\ \bibnamefont {Kim}}, \bibinfo
  {author} {\bibfnamefont {G.~H.}\ \bibnamefont {Lee}}, \bibinfo {author}
  {\bibfnamefont {Y.~S.}\ \bibnamefont {Lee}}, \bibinfo {author} {\bibfnamefont
  {J.~Y.}\ \bibnamefont {Park}}, \bibinfo {author} {\bibfnamefont {D.~J.}\
  \bibnamefont {Cho}}, \ and\ \bibinfo {author} {\bibfnamefont {C.~H.}\
  \bibnamefont {Nam}},\ }\href {\doibase 10.1103/PhysRevLett.94.243901}
  {\bibfield  {journal} {\bibinfo  {journal} {Physical Review Letters}\
  }\textbf {\bibinfo {volume} {94}},\ \bibinfo {pages} {2} (\bibinfo {year}
  {2005})}\BibitemShut {NoStop}%
\bibitem [{\citenamefont {{Roscam Abbing}}\ \emph {et~al.}(2020)\citenamefont
  {{Roscam Abbing}}, \citenamefont {Campi}, \citenamefont {Sajjadian},
  \citenamefont {Lin}, \citenamefont {Smorenburg},\ and\ \citenamefont
  {Kraus}}]{RoscamAbbing2019}%
  \BibitemOpen
  \bibfield  {author} {\bibinfo {author} {\bibfnamefont {S.}~\bibnamefont
  {{Roscam Abbing}}}, \bibinfo {author} {\bibfnamefont {F.}~\bibnamefont
  {Campi}}, \bibinfo {author} {\bibfnamefont {F.~S.}\ \bibnamefont
  {Sajjadian}}, \bibinfo {author} {\bibfnamefont {N.}~\bibnamefont {Lin}},
  \bibinfo {author} {\bibfnamefont {P.}~\bibnamefont {Smorenburg}}, \ and\
  \bibinfo {author} {\bibfnamefont {P.~M.}\ \bibnamefont {Kraus}},\ }\href
  {\doibase 10.1103/PhysRevApplied.13.054029} {\bibfield  {journal} {\bibinfo
  {journal} {Physical Review Applied}\ }\textbf {\bibinfo {volume} {13}},\
  \bibinfo {pages} {1} (\bibinfo {year} {2020})}\ \BibitemShut {NoStop}%
\bibitem [{\citenamefont {Yamanoi}\ \emph {et~al.}(2010)\citenamefont
  {Yamanoi}, \citenamefont {Sakai}, \citenamefont {Nakazato}, \citenamefont
  {Estacio}, \citenamefont {Shimizu}, \citenamefont {Sarukura}, \citenamefont
  {Ehrentraut}, \citenamefont {Fukuda}, \citenamefont {Nagasono}, \citenamefont
  {Togashi}, \citenamefont {Matsubara}, \citenamefont {Tono}, \citenamefont
  {Yabashi}, \citenamefont {Kimura}, \citenamefont {Ohashi},\ and\
  \citenamefont {Ishikawa}}]{Yamanoi2010}%
  \BibitemOpen
  \bibfield  {author} {\bibinfo {author} {\bibfnamefont {K.}~\bibnamefont
  {Yamanoi}}, \bibinfo {author} {\bibfnamefont {K.}~\bibnamefont {Sakai}},
  \bibinfo {author} {\bibfnamefont {T.}~\bibnamefont {Nakazato}}, \bibinfo
  {author} {\bibfnamefont {E.}~\bibnamefont {Estacio}}, \bibinfo {author}
  {\bibfnamefont {T.}~\bibnamefont {Shimizu}}, \bibinfo {author} {\bibfnamefont
  {N.}~\bibnamefont {Sarukura}}, \bibinfo {author} {\bibfnamefont
  {D.}~\bibnamefont {Ehrentraut}}, \bibinfo {author} {\bibfnamefont
  {T.}~\bibnamefont {Fukuda}}, \bibinfo {author} {\bibfnamefont
  {M.}~\bibnamefont {Nagasono}}, \bibinfo {author} {\bibfnamefont
  {T.}~\bibnamefont {Togashi}}, \bibinfo {author} {\bibfnamefont
  {S.}~\bibnamefont {Matsubara}}, \bibinfo {author} {\bibfnamefont
  {K.}~\bibnamefont {Tono}}, \bibinfo {author} {\bibfnamefont {M.}~\bibnamefont
  {Yabashi}}, \bibinfo {author} {\bibfnamefont {H.}~\bibnamefont {Kimura}},
  \bibinfo {author} {\bibfnamefont {H.}~\bibnamefont {Ohashi}}, \ and\ \bibinfo
  {author} {\bibfnamefont {T.}~\bibnamefont {Ishikawa}},\ }in\ \href {\doibase
  10.1016/j.optmat.2010.04.039} {\emph {\bibinfo {booktitle} {Optical
  Materials}}},\ Vol.~\bibinfo {volume} {32}\ (\bibinfo  {publisher} {Elsevier
  B.V.},\ \bibinfo {year} {2010})\ pp.\ \bibinfo {pages}
  {1305--1308}\BibitemShut {NoStop}%
\bibitem [{\citenamefont {Manschwetus}\ \emph {et~al.}(2016)\citenamefont
  {Manschwetus}, \citenamefont {Rading}, \citenamefont {Campi}, \citenamefont
  {Maclot}, \citenamefont {Coudert-Alteirac}, \citenamefont {Lahl},
  \citenamefont {Wikmark}, \citenamefont {Rudawski}, \citenamefont {Heyl},
  \citenamefont {Farkas}, \citenamefont {Mohamed}, \citenamefont {L'Huillier},\
  and\ \citenamefont {Johnsson}}]{Manschwetus2016}%
  \BibitemOpen
  \bibfield  {author} {\bibinfo {author} {\bibfnamefont {B.}~\bibnamefont
  {Manschwetus}}, \bibinfo {author} {\bibfnamefont {L.}~\bibnamefont {Rading}},
  \bibinfo {author} {\bibfnamefont {F.}~\bibnamefont {Campi}}, \bibinfo
  {author} {\bibfnamefont {S.}~\bibnamefont {Maclot}}, \bibinfo {author}
  {\bibfnamefont {H.}~\bibnamefont {Coudert-Alteirac}}, \bibinfo {author}
  {\bibfnamefont {J.}~\bibnamefont {Lahl}}, \bibinfo {author} {\bibfnamefont
  {H.}~\bibnamefont {Wikmark}}, \bibinfo {author} {\bibfnamefont
  {P.}~\bibnamefont {Rudawski}}, \bibinfo {author} {\bibfnamefont {C.~M.}\
  \bibnamefont {Heyl}}, \bibinfo {author} {\bibfnamefont {B.}~\bibnamefont
  {Farkas}}, \bibinfo {author} {\bibfnamefont {T.}~\bibnamefont {Mohamed}},
  \bibinfo {author} {\bibfnamefont {A.}~\bibnamefont {L'Huillier}}, \ and\
  \bibinfo {author} {\bibfnamefont {P.}~\bibnamefont {Johnsson}},\ }\href
  {\doibase 10.1103/PhysRevA.93.061402} {\bibfield  {journal} {\bibinfo
  {journal} {Physical Review A}\ }\textbf {\bibinfo {volume} {93}} (\bibinfo
  {year} {2016}),\ 10.1103/PhysRevA.93.061402}\BibitemShut {NoStop}%
\bibitem [{\citenamefont {Campbell}\ \emph {et~al.}(1999)\citenamefont
  {Campbell}, \citenamefont {Kalia}, \citenamefont {Nakano}, \citenamefont
  {Vashishta}, \citenamefont {Ogata},\ and\ \citenamefont
  {Rodgers}}]{Campbell1999}%
  \BibitemOpen
  \bibfield  {author} {\bibinfo {author} {\bibfnamefont {T.}~\bibnamefont
  {Campbell}}, \bibinfo {author} {\bibfnamefont {R.~K.}\ \bibnamefont {Kalia}},
  \bibinfo {author} {\bibfnamefont {A.}~\bibnamefont {Nakano}}, \bibinfo
  {author} {\bibfnamefont {P.}~\bibnamefont {Vashishta}}, \bibinfo {author}
  {\bibfnamefont {S.}~\bibnamefont {Ogata}}, \ and\ \bibinfo {author}
  {\bibfnamefont {S.}~\bibnamefont {Rodgers}},\ }\href {\doibase
  10.1103/PhysRevLett.82.4866} {\bibfield  {journal} {\bibinfo  {journal}
  {Physical Review Letters}\ }\textbf {\bibinfo {volume} {82}},\ \bibinfo
  {pages} {4866} (\bibinfo {year} {1999})}\BibitemShut {NoStop}%
\bibitem [{\citenamefont {Frassetto}\ \emph {et~al.}(2010)\citenamefont
  {Frassetto}, \citenamefont {Coraggia}, \citenamefont {Poletto}, \citenamefont
  {Guerassimova}, \citenamefont {Dziarzhytski}, \citenamefont {Ploenjes},\ and\
  \citenamefont {Weigelt}}]{Frassetto2010}%
  \BibitemOpen
  \bibfield  {author} {\bibinfo {author} {\bibfnamefont {F.}~\bibnamefont
  {Frassetto}}, \bibinfo {author} {\bibfnamefont {S.}~\bibnamefont {Coraggia}},
  \bibinfo {author} {\bibfnamefont {L.}~\bibnamefont {Poletto}}, \bibinfo
  {author} {\bibfnamefont {N.}~\bibnamefont {Guerassimova}}, \bibinfo {author}
  {\bibfnamefont {S.}~\bibnamefont {Dziarzhytski}}, \bibinfo {author}
  {\bibfnamefont {E.}~\bibnamefont {Ploenjes}}, \ and\ \bibinfo {author}
  {\bibfnamefont {H.}~\bibnamefont {Weigelt}},\ }in\ \href {\doibase
  10.1117/12.861615} {\emph {\bibinfo {booktitle} {Advances in X-Ray/EUV Optics
  and Components V}}},\ Vol.\ \bibinfo {volume} {7802}\ (\bibinfo  {publisher}
  {SPIE},\ \bibinfo {year} {2010})\ p.\ \bibinfo {pages} {780209}\BibitemShut
  {NoStop}%
\bibitem [{\citenamefont {Scholze}, \citenamefont {Rabus},\ and\ \citenamefont
  {Ulm}(1998)}]{Scholze1998}%
  \BibitemOpen
  \bibfield  {author} {\bibinfo {author} {\bibfnamefont {F.}~\bibnamefont
  {Scholze}}, \bibinfo {author} {\bibfnamefont {H.}~\bibnamefont {Rabus}}, \
  and\ \bibinfo {author} {\bibfnamefont {G.}~\bibnamefont {Ulm}},\ }\href
  {\doibase 10.1063/1.368398} {\bibfield  {journal} {\bibinfo  {journal}
  {Journal of Applied Physics}\ }\textbf {\bibinfo {volume} {84}},\ \bibinfo
  {pages} {2926} (\bibinfo {year} {1998})}\BibitemShut {NoStop}%
\bibitem [{\citenamefont {Wang}\ \emph {et~al.}(2013)\citenamefont {Wang},
  \citenamefont {Chini}, \citenamefont {Cheng}, \citenamefont {Wu},\ and\
  \citenamefont {Chang}}]{Wang2013}%
  \BibitemOpen
  \bibfield  {author} {\bibinfo {author} {\bibfnamefont {X.}~\bibnamefont
  {Wang}}, \bibinfo {author} {\bibfnamefont {M.}~\bibnamefont {Chini}},
  \bibinfo {author} {\bibfnamefont {Y.}~\bibnamefont {Cheng}}, \bibinfo
  {author} {\bibfnamefont {Y.}~\bibnamefont {Wu}}, \ and\ \bibinfo {author}
  {\bibfnamefont {Z.}~\bibnamefont {Chang}},\ }\href {\doibase
  10.1364/AO.52.000323} {\bibfield  {journal} {\bibinfo  {journal} {Applied
  Optics}\ }\textbf {\bibinfo {volume} {52}},\ \bibinfo {pages} {323} (\bibinfo
  {year} {2013})}\BibitemShut {NoStop}%
\bibitem [{\citenamefont {Madden}, \citenamefont {Ederer},\ and\ \citenamefont
  {Codling}(1969)}]{Madden1969}%
  \BibitemOpen
  \bibfield  {author} {\bibinfo {author} {\bibfnamefont {R.~P.}\ \bibnamefont
  {Madden}}, \bibinfo {author} {\bibfnamefont {D.~L.}\ \bibnamefont {Ederer}},
  \ and\ \bibinfo {author} {\bibfnamefont {K.}~\bibnamefont {Codling}},\ }\href
  {\doibase 10.1103/PhysRev.177.136} {\bibfield  {journal} {\bibinfo  {journal}
  {Physical Review}\ }\textbf {\bibinfo {volume} {177}},\ \bibinfo {pages}
  {136} (\bibinfo {year} {1969})}\BibitemShut {NoStop}%
\bibitem [{\citenamefont {Wang}\ \emph {et~al.}(2010)\citenamefont {Wang},
  \citenamefont {Chini}, \citenamefont {Chen}, \citenamefont {Zhang},
  \citenamefont {He}, \citenamefont {Cheng}, \citenamefont {Wu}, \citenamefont
  {Thumm},\ and\ \citenamefont {Chang}}]{Wang2010}%
  \BibitemOpen
  \bibfield  {author} {\bibinfo {author} {\bibfnamefont {H.}~\bibnamefont
  {Wang}}, \bibinfo {author} {\bibfnamefont {M.}~\bibnamefont {Chini}},
  \bibinfo {author} {\bibfnamefont {S.}~\bibnamefont {Chen}}, \bibinfo {author}
  {\bibfnamefont {C.~H.}\ \bibnamefont {Zhang}}, \bibinfo {author}
  {\bibfnamefont {F.}~\bibnamefont {He}}, \bibinfo {author} {\bibfnamefont
  {Y.}~\bibnamefont {Cheng}}, \bibinfo {author} {\bibfnamefont
  {Y.}~\bibnamefont {Wu}}, \bibinfo {author} {\bibfnamefont {U.}~\bibnamefont
  {Thumm}}, \ and\ \bibinfo {author} {\bibfnamefont {Z.}~\bibnamefont
  {Chang}},\ }\href {\doibase 10.1103/PhysRevLett.105.143002} {\bibfield
  {journal} {\bibinfo  {journal} {Physical Review Letters}\ }\textbf {\bibinfo
  {volume} {105}},\ \bibinfo {pages} {3} (\bibinfo {year} {2010})}\BibitemShut
  {NoStop}%
\bibitem [{\citenamefont {Codling}, \citenamefont {Madden},\ and\ \citenamefont
  {Ederer}(1967)}]{Codling1967}%
  \BibitemOpen
  \bibfield  {author} {\bibinfo {author} {\bibfnamefont {K.}~\bibnamefont
  {Codling}}, \bibinfo {author} {\bibfnamefont {R.~P.}\ \bibnamefont {Madden}},
  \ and\ \bibinfo {author} {\bibfnamefont {D.~L.}\ \bibnamefont {Ederer}},\
  }\href {\doibase 10.1103/PhysRev.155.26} {\bibfield  {journal} {\bibinfo
  {journal} {Physical Review}\ }\textbf {\bibinfo {volume} {155}},\ \bibinfo
  {pages} {26} (\bibinfo {year} {1967})}\BibitemShut {NoStop}%
\bibitem [{\citenamefont {Samson}\ and\ \citenamefont
  {Haddad}(1974)}]{Samson1974}%
  \BibitemOpen
  \bibfield  {author} {\bibinfo {author} {\bibfnamefont {J.~A.}\ \bibnamefont
  {Samson}}\ and\ \bibinfo {author} {\bibfnamefont {G.~N.}\ \bibnamefont
  {Haddad}},\ }\href {\doibase 10.1364/JOSA.64.001346} {\bibfield  {journal}
  {\bibinfo  {journal} {J Opt Soc Am}\ }\textbf {\bibinfo {volume} {64}},\
  \bibinfo {pages} {1346} (\bibinfo {year} {1974})}\BibitemShut {NoStop}%
\bibitem [{\citenamefont {Moine}\ \emph {et~al.}(2007)\citenamefont {Moine},
  \citenamefont {Bizarri}, \citenamefont {Varrel},\ and\ \citenamefont
  {Rivoire}}]{Moine2007}%
  \BibitemOpen
  \bibfield  {author} {\bibinfo {author} {\bibfnamefont {B.}~\bibnamefont
  {Moine}}, \bibinfo {author} {\bibfnamefont {G.}~\bibnamefont {Bizarri}},
  \bibinfo {author} {\bibfnamefont {B.}~\bibnamefont {Varrel}}, \ and\ \bibinfo
  {author} {\bibfnamefont {J.~Y.}\ \bibnamefont {Rivoire}},\ }\href {\doibase
  10.1016/j.optmat.2006.05.004} {\bibfield  {journal} {\bibinfo  {journal}
  {Optical Materials}\ }\textbf {\bibinfo {volume} {29}},\ \bibinfo {pages}
  {1148} (\bibinfo {year} {2007})}\BibitemShut {NoStop}%
\bibitem [{\citenamefont {Sigmond}(1966)}]{Sigmond1966}%
  \BibitemOpen
  \bibfield  {author} {\bibinfo {author} {\bibfnamefont {R.~S.}\ \bibnamefont
  {Sigmond}},\ }\href {\doibase 10.1088/0508-3443/17/10/308} {\bibfield
  {journal} {\bibinfo  {journal} {British Journal of Applied Physics}\ }\textbf
  {\bibinfo {volume} {17}},\ \bibinfo {pages} {1307} (\bibinfo {year}
  {1966})}\BibitemShut {NoStop}%
\bibitem [{\citenamefont {Waynant}\ and\ \citenamefont
  {Elton}(1971)}]{Waynant1971}%
  \BibitemOpen
  \bibfield  {author} {\bibinfo {author} {\bibfnamefont {R.}~\bibnamefont
  {Waynant}}\ and\ \bibinfo {author} {\bibfnamefont {R.}~\bibnamefont
  {Elton}},\ }\href {\doibase 10.1016/b978-0-12-356250-0.50037-3} {\emph
  {\bibinfo {title} {Organic Scintillators and Scintillation Counting}}}\
  (\bibinfo  {publisher} {ACADEMIC PRESS, INC.},\ \bibinfo {year} {1971})\ pp.\
  \bibinfo {pages} {467--477}\BibitemShut {NoStop}%
\bibitem [{\citenamefont {Baker}, \citenamefont {Brocklehurst},\ and\
  \citenamefont {Holton}(1987)}]{Baker1987}%
  \BibitemOpen
  \bibfield  {author} {\bibinfo {author} {\bibfnamefont {G.~J.}\ \bibnamefont
  {Baker}}, \bibinfo {author} {\bibfnamefont {B.}~\bibnamefont {Brocklehurst}},
  \ and\ \bibinfo {author} {\bibfnamefont {I.~R.}\ \bibnamefont {Holton}},\
  }\href {\doibase 10.1088/0022-3700/20/10/003} {\bibfield  {journal} {\bibinfo
   {journal} {Journal of Physics B: Atomic and Molecular Physics}\ }\textbf
  {\bibinfo {volume} {20}},\ \bibinfo {pages} {L305} (\bibinfo {year}
  {1987})}\BibitemShut {NoStop}%
\bibitem [{\citenamefont {Riedel}\ \emph {et~al.}(2001)\citenamefont {Riedel},
  \citenamefont {Hernandez-Pozos}, \citenamefont {Palmer}, \citenamefont
  {Baggott}, \citenamefont {Kolasinski},\ and\ \citenamefont
  {Foord}}]{Kolasinski2002}%
  \BibitemOpen
  \bibfield  {author} {\bibinfo {author} {\bibfnamefont {D.}~\bibnamefont
  {Riedel}}, \bibinfo {author} {\bibfnamefont {J.~L.}\ \bibnamefont
  {Hernandez-Pozos}}, \bibinfo {author} {\bibfnamefont {R.~E.}\ \bibnamefont
  {Palmer}}, \bibinfo {author} {\bibfnamefont {S.}~\bibnamefont {Baggott}},
  \bibinfo {author} {\bibfnamefont {K.~W.}\ \bibnamefont {Kolasinski}}, \ and\
  \bibinfo {author} {\bibfnamefont {J.~S.}\ \bibnamefont {Foord}},\ }\href
  {\doibase 10.1063/1.1351835} {\bibfield  {journal} {\bibinfo  {journal}
  {Review of Scientific Instruments}\ }\textbf {\bibinfo {volume} {72}},\
  \bibinfo {pages} {1977} (\bibinfo {year} {2001})}\BibitemShut {NoStop}%
\bibitem [{\citenamefont {Brocklehurst}(1997)}]{Brocklehurst1997}%
  \BibitemOpen
  \bibfield  {author} {\bibinfo {author} {\bibfnamefont {B.}~\bibnamefont
  {Brocklehurst}},\ }\href {\doibase 10.1016/S0969-806X(97)00029-7} {\bibfield
  {journal} {\bibinfo  {journal} {Radiation Physics and Chemistry}\ }\textbf
  {\bibinfo {volume} {50}},\ \bibinfo {pages} {213} (\bibinfo {year}
  {1997})}\BibitemShut {NoStop}%
\bibitem [{\citenamefont {Brocklehurst}\ and\ \citenamefont
  {Munro}(2008)}]{Brocklehurst1992a}%
  \BibitemOpen
  \bibfield  {author} {\bibinfo {author} {\bibfnamefont {B.}~\bibnamefont
  {Brocklehurst}}\ and\ \bibinfo {author} {\bibfnamefont {I.~H.}\ \bibnamefont
  {Munro}},\ }in\ \href {\doibase 10.1063/1.42504} {\emph {\bibinfo {booktitle}
  {AIP Conference Proceedings}}},\ Vol.\ \bibinfo {volume} {258}\ (\bibinfo
  {year} {2008})\ pp.\ \bibinfo {pages} {465--473}\BibitemShut {NoStop}%
\bibitem [{\citenamefont {Xue}\ \emph {et~al.}(2015)\citenamefont {Xue},
  \citenamefont {Sun}, \citenamefont {Chen}, \citenamefont {Gong},
  \citenamefont {Yao}, \citenamefont {Zhang}, \citenamefont {Qian},\ and\
  \citenamefont {Lu}}]{Xue2015}%
  \BibitemOpen
  \bibfield  {author} {\bibinfo {author} {\bibfnamefont {P.}~\bibnamefont
  {Xue}}, \bibinfo {author} {\bibfnamefont {J.}~\bibnamefont {Sun}}, \bibinfo
  {author} {\bibfnamefont {P.}~\bibnamefont {Chen}}, \bibinfo {author}
  {\bibfnamefont {P.}~\bibnamefont {Gong}}, \bibinfo {author} {\bibfnamefont
  {B.}~\bibnamefont {Yao}}, \bibinfo {author} {\bibfnamefont {Z.}~\bibnamefont
  {Zhang}}, \bibinfo {author} {\bibfnamefont {C.}~\bibnamefont {Qian}}, \ and\
  \bibinfo {author} {\bibfnamefont {R.}~\bibnamefont {Lu}},\ }\href {\doibase
  10.1039/c5tc00267b} {\bibfield  {journal} {\bibinfo  {journal} {Journal of
  Materials Chemistry C}\ }\textbf {\bibinfo {volume} {3}},\ \bibinfo {pages}
  {4086} (\bibinfo {year} {2015})}\BibitemShut {NoStop}%
\bibitem [{\citenamefont {Wang}\ \emph {et~al.}(2016)\citenamefont {Wang},
  \citenamefont {Zhang}, \citenamefont {Zhang}, \citenamefont {Wang},
  \citenamefont {Wu}, \citenamefont {Liang}, \citenamefont {Hao}, \citenamefont
  {Fu}, \citenamefont {Zhao},\ and\ \citenamefont {Zhang}}]{Wang2016}%
  \BibitemOpen
  \bibfield  {author} {\bibinfo {author} {\bibfnamefont {Y.}~\bibnamefont
  {Wang}}, \bibinfo {author} {\bibfnamefont {G.}~\bibnamefont {Zhang}},
  \bibinfo {author} {\bibfnamefont {W.}~\bibnamefont {Zhang}}, \bibinfo
  {author} {\bibfnamefont {X.}~\bibnamefont {Wang}}, \bibinfo {author}
  {\bibfnamefont {Y.}~\bibnamefont {Wu}}, \bibinfo {author} {\bibfnamefont
  {T.}~\bibnamefont {Liang}}, \bibinfo {author} {\bibfnamefont
  {X.}~\bibnamefont {Hao}}, \bibinfo {author} {\bibfnamefont {H.}~\bibnamefont
  {Fu}}, \bibinfo {author} {\bibfnamefont {Y.}~\bibnamefont {Zhao}}, \ and\
  \bibinfo {author} {\bibfnamefont {D.}~\bibnamefont {Zhang}},\ }\href
  {\doibase 10.1002/smll.201601516} {\bibfield  {journal} {\bibinfo  {journal}
  {Small}\ }\textbf {\bibinfo {volume} {12}},\ \bibinfo {pages} {6554}
  (\bibinfo {year} {2016})}\BibitemShut {NoStop}%
\bibitem [{\citenamefont {Liu}\ \emph {et~al.}(2020)\citenamefont {Liu},
  \citenamefont {Li}, \citenamefont {Ma}, \citenamefont {Xu}, \citenamefont
  {Ma},\ and\ \citenamefont {Jia}}]{Liu2020}%
  \BibitemOpen
  \bibfield  {author} {\bibinfo {author} {\bibfnamefont {Y.}~\bibnamefont
  {Liu}}, \bibinfo {author} {\bibfnamefont {A.}~\bibnamefont {Li}}, \bibinfo
  {author} {\bibfnamefont {Z.}~\bibnamefont {Ma}}, \bibinfo {author}
  {\bibfnamefont {W.}~\bibnamefont {Xu}}, \bibinfo {author} {\bibfnamefont
  {Z.}~\bibnamefont {Ma}}, \ and\ \bibinfo {author} {\bibfnamefont
  {X.}~\bibnamefont {Jia}},\ }\href {\doibase 10.1039/d0cp02783a} {\bibfield
  {journal} {\bibinfo  {journal} {Physical Chemistry Chemical Physics}\
  }\textbf {\bibinfo {volume} {22}},\ \bibinfo {pages} {19195} (\bibinfo {year}
  {2020})}\BibitemShut {NoStop}%
\bibitem [{\citenamefont {Wu}\ \emph {et~al.}(2020)\citenamefont {Wu},
  \citenamefont {Bespalov}, \citenamefont {Witte}, \citenamefont {Lugier},
  \citenamefont {Haitjema}, \citenamefont {Vockenhuber}, \citenamefont
  {Ekinci}, \citenamefont {Watts}, \citenamefont {Brouwer},\ and\ \citenamefont
  {Castellanos}}]{Wu2020}%
  \BibitemOpen
  \bibfield  {author} {\bibinfo {author} {\bibfnamefont {L.}~\bibnamefont
  {Wu}}, \bibinfo {author} {\bibfnamefont {I.}~\bibnamefont {Bespalov}},
  \bibinfo {author} {\bibfnamefont {K.}~\bibnamefont {Witte}}, \bibinfo
  {author} {\bibfnamefont {O.}~\bibnamefont {Lugier}}, \bibinfo {author}
  {\bibfnamefont {J.}~\bibnamefont {Haitjema}}, \bibinfo {author}
  {\bibfnamefont {M.}~\bibnamefont {Vockenhuber}}, \bibinfo {author}
  {\bibfnamefont {Y.}~\bibnamefont {Ekinci}}, \bibinfo {author} {\bibfnamefont
  {B.}~\bibnamefont {Watts}}, \bibinfo {author} {\bibfnamefont {A.~M.}\
  \bibnamefont {Brouwer}}, \ and\ \bibinfo {author} {\bibfnamefont
  {S.}~\bibnamefont {Castellanos}},\ }\href {\doibase 10.1039/d0tc03216f}
  {\bibfield  {journal} {\bibinfo  {journal} {Journal of Materials Chemistry
  C}\ }\textbf {\bibinfo {volume} {8}},\ \bibinfo {pages} {14757} (\bibinfo
  {year} {2020})}\BibitemShut {NoStop}%
\bibitem [{\citenamefont {Henke}, \citenamefont {Gullikson},\ and\
  \citenamefont {Davis}(1993)}]{Henke1993}%
  \BibitemOpen
  \bibfield  {author} {\bibinfo {author} {\bibfnamefont {B.~L.}\ \bibnamefont
  {Henke}}, \bibinfo {author} {\bibfnamefont {E.~M.}\ \bibnamefont
  {Gullikson}}, \ and\ \bibinfo {author} {\bibfnamefont {J.~C.}\ \bibnamefont
  {Davis}},\ }\href {\doibase 10.1006/adnd.1993.1013} {\bibfield  {journal}
  {\bibinfo  {journal} {Atomic Data and Nuclear Data Tables}\ }\textbf
  {\bibinfo {volume} {54}},\ \bibinfo {pages} {181} (\bibinfo {year}
  {1993})}\BibitemShut {NoStop}%
\end{thebibliography}
\end{document}